\documentclass[prx,twocolumn,showpacs,showkeys,groupedaddress,longbibliography,preprintnumbers]{revtex4-1}

\usepackage{verbatim}
\usepackage{longtable} 
\usepackage{subfigure}
\usepackage{amsmath}
\usepackage{wrapfig}
\usepackage{epsfig}
\usepackage{float}
\usepackage{amsmath,amssymb,bm}
\usepackage{array}
\usepackage{color}
\usepackage{bbold,ulem}
\renewcommand{\emph}{\textit}

\usepackage{tikz}

\renewcommand{\vec}[1]{\mathbf{#1}}

\newcommand{\SNAKE}{\textit{SNAKE}}

\usepackage{graphicx}

\begin{document}

\title{Numerical treatment of the Boltzmann equation for self-propelled particle systems
}

\preprint{LMU-ASC 68/14}

\author{Florian Th\"uroff}
\author{Christoph A. Weber} 
\author{Erwin Frey}

\affiliation{Arnold Sommerfeld Center for Theoretical Physics and Center for NanoScience, Department of Physics, Ludwig-Maximilians-University, Munich, Germany}

\date{\today}

\begin{abstract}
Kinetic theories constitute one of the most promising tools to decipher the characteristic spatio-temporal dynamics in systems of actively propelled particles. In this context, the Boltzmann equation plays a pivotal role, since it provides a natural translation between a particle-level description of the system's dynamics and the corresponding hydrodynamic fields. Yet, the intricate mathematical structure of the Boltzmann equation substantially limits the progress toward a full understanding of this equation by solely analytical means. Here, we propose a
general framework to numerically solve the Boltzmann equation for self-propelled particle systems in two spatial dimensions and with arbitrary boundary conditions. 
We discuss potential applications of this numerical framework to active matter systems, and use the algorithm to give a detailed analysis to a model system of self-propelled particles with polar interactions.
In accordance with previous studies, we find that spatially homogeneous isotropic and broken symmetry states populate two distinct regions in parameter space, which are separated by a narrow region of spatially inhomogeneous, density-segregated moving patterns. We find clear evidence that these three regions in parameter space are connected by first order phase transitions, 
and that the transition between the spatially homogeneous isotropic and polar ordered phases bears striking similarities to liquid-gas phase transitions in equilibrium systems.
Within the density segregated parameter regime, we find a novel stable limit-cycle solution of the Boltzmann equation, which consists of parallel lanes of polar clusters moving in opposite directions, so as to render the overall symmetry of the system's ordered state nematic, despite purely polar interactions on the level of single particles. 
\end{abstract}


\maketitle

\vspace{1cm}

\section{\label{sec:Introduction}Introduction}

Developing a deeper understanding of active matter \cite{RamaswamyReview_with_John,Ramaswamy_Review,Vicsek_Review,Marchetti:2012ws} has been the major focus of a considerable amount of theoretical work over the last decades \cite{Vicsek,Toner_Tu_1995,Toner_Tu_1998,Chate_2004,Peruani_rods,2006PhRvL..97i0602M,Marchetti2008,Baskaran_Marchetti_2008,Chate_Variations,Chate_long,Ginelli,Mishra_Marchetti_2010,Peruani_Traffic_Jams,Gopinath:2012wq,Farrell:2012vf,2013PhRvL.111g8101S}. In recent years, kinetic theory has gained considerable popularity to assess the ordering behavior in systems of actively propelled particles~\cite{Aronson_MT,Bertin_short,Ben-Naim_2006,Baskaran:2008he,Baskaran:2008bl,Bertin_long,Aranson_Bakterien,Ihle_2011,Chou:2012wq,Peshkov:2012uu,Peshkov:2012tu,Barath_bac_susp_2013,Weber_NJP_2013,Thuroff_2013,1367-2630-15-8-085032,1367-2630-16-3-035003}. In this context, the Boltzmann equation provides a particularly compelling approach to active matter systems. Apart from its inherent limitations, which are largely due the assumptions of binary particle interactions and molecular chaos (cf. Refs.~\cite{Ihle_2011,Thuroff_2013,Hanke_2013}), the following advantages of this framework are manifest: (i) The structure of the Boltzmann equation, relating convection and collision processes on the level of the one-particle distribution function, is ideally suited to explicitly implement a microscopic picture of particle dynamics. (ii) Due to its mesoscopic character, the Boltzmann equation provides an immediate connection to the system's hydrodynamic variables, which naturally arise in the form of the various moments of the one-particle distribution function. Therefore, the Boltzmann equation sets up a direct link between the microscopic dynamics of the system's constituent particles and the corresponding physics emerging on hydrodynamic length and time scales. Moreover, since the Boltzmann equation keeps track of all parameters used to formulate a particle-level description of the system under study, the resulting hydrodynamics can be studied explicitly in terms of those microscopic parameters.

The Boltzmann equation and related kinetic frameworks have been employed in a number of previous studies on two-dimensional systems of self-propelled particles \cite{Bertin_short,Aranson_Bakterien,Bertin_long,Ihle_2011,Farrell:2012vf,Peshkov:2012uu,Peshkov:2012tu,Weber_NJP_2013,Ihle:2013jg}, mainly to derive a hydrodynamic description. To this end, an expansion technique is used to derive an infinite hierarchy of coupled equations of motion for the Fourier modes of the angular degrees of freedom of the one-particle distribution function. 
Since the zeroth, first, second, and higher order Fourier modes are directly related to the local average particle number, polar, nematic, and higher symmetry order parameters, the Fourier space description lends itself as a starting point to extract a hydrodynamic description. In general, this is achieved by truncating the hierarchy of Fourier space equations by means of a perturbation ansatz for the Fourier modes, which itself typically depends on system's symmetries.
Following a similar strategy, a numerical approach based on an Enskog-like kinetic theory has recently been proposed in Ref. \cite{Ihle:2013jg}.

While such Fourier space approaches have been successfully applied to investigate the onset of collective motion, the validity of the underlying truncation schemes remains largely elusive for parameters deeper inside the ordered region.
In this work, we propose a Boltzmann equation based approach which we will refer to as \SNAKE~(\underline{s}olving \underline{n}umerically \underline{a}ctive \underline{k}inetic \underline{e}quations). \SNAKE~focuses on a direct numerical solution of the Boltzmann equation in real space, without any explicit recourse to the aforementioned coupled hierarchy of equations in Fourier space.
Such a direct numerical approach offers a number of advantages in the study of active systems, which can be summarized as follows. First of all, the real space Boltzmann equation constitutes a closed description for the dynamics of the one-particle distribution function.
Consequently, the range of validity of this direct numerical strategy is restricted only by inherent limitations of the Boltzmann equation itself but is not principally confined to a parameter region around the onset of collective motion. 
Secondly, \SNAKE~provides full access to the flexibility of the Boltzmann equation in the implementation of specific model systems which are based on a concrete microscopic picture of particle dynamics. This, thirdly, equips this approach with particularly powerful capabilities to incorporate physical boundary conditions, which can be formulated on the basis of actual particle-wall interactions rather than on the level of macroscopic, hydrodynamic field variables. Finally, since all pertinent hydrodynamic fields are easily computed from the one-particle distribution function by taking appropriate averages, this numerical approach allows to fully exploit the Boltzmann equation's potential to mediate between the physics on microscopic and macroscopic scales.

To demonstrate the benefits of \SNAKE, we will present a detailed study of an archetypical model for active particles subject to binary, polar particle interactions, which has been proposed originally by Bertin et al. \cite{Bertin_short,Bertin_long} and which we will henceforth refer to as ``BDG model''. As has been established previously \cite{Bertin_long}, the BDG model transitions from a spatially homogeneous, fully isotropic state at low densities / high noise levels to a collectively moving state at high densities / low noise levels. In the vicinity of this phase transition, the polar order parameter field was shown to be susceptible to a longitudinal instability \cite{Bertin_long}, which in turn has been linked to the emergence of stable solitary wave patterns propagating on an otherwise isotropic low density background \cite{Ihle:2013jg,Thuroff_2013}. Well beyond the transition line to collective motion, this longitudinal instability disappears from the Boltzmann equation \cite{Bertin_long} and a spatially homogeneous state of collective motion has been shown to emerge \cite{Thuroff_2013}.

Our work is structured as follows. In section \ref{sec:Discretization}, we discuss a discretization scheme for the Boltzmann equation of self-propelled particles in two spatial dimensions. The central result of this section is an update rule, Eq.~\eqref{eq:UpdateRule}, which lends itself to direct numerical implementation. This section also contains a discussion of the general strategy to implement virtually arbitrary boundary conditions and geometries.
Readers not interested in the details of the actual discretization procedure underlying the \SNAKE~algorithm may skip sections \ref{sec:DimlessBE} -- \ref{sec:BCs} and proceed to section \ref{sec:Applications}.

There, we will exploit \SNAKE's capabilities to address the stationary states of the BDG model over a broad range in parameter space (section \ref{sec:PhaseDiagram}). In particular, we will show that our implementation of \SNAKE~consistently reproduces previously established analytical results on this model \cite{Bertin_long}. In section \ref{sec:LD_transition} we focus on the onset of collective motion, which is accompanied by the formation of density segregated patterns and hysteresis effects, as previously reported in numerical studies of the Vicsek model \cite{Chate_2004,Chate_long, Ihle:2013jg}. Additionally, in section \ref{sec:DSP}, we extend previous results on the flocking transition in polar systems \cite{Bertin_long, Ihle_2011, Ihle:2013jg} well beyond the onset of order. 
Specifically, we demonstrate the presence of an additional first order phase transition from a spatially inhomogeneous, density segregated phase to a spatially homogeneous polar phase, taking place well inside the ordered, broken-symmetry parameter regime (section \ref{sec:HD_transition}). Overall, we find striking similarities between the flocking transition to spatially homogeneous polar order, on the one hand, and the features of liquid-gas phase transitions in equilibrium systems, on the other hand. Analogous observations have been made previously in the context of a lattice model of active Ising spins, where a discrete up-down symmetry is spontaneously broken in two spatial dimensions \cite{2013PhRvL.111g8101S}.
In section \ref{sec:Cluster}, we shift our focus to a deeper discussion of density segregated patterns and report on the emergence of previously unseen ``cluster-lane patterns'' which seem to occupy the same parameter region as the familiar solitary wave patterns and to constitute a stable limit-cycle solution of the underlying Boltzmann equation. These patterns consist of parallel lanes of polar clusters moving in opposite directions; cf. Figs.~\ref{fig:cluster_particle_currents}(b,c) and Figs.~\ref{fig:cluster_lane_geometry}(b,c) for snapshots. Intriguingly, the stability of such patterns provides a mechanism to establish states of nematic macroscopic order, arising from perfectly polar particle interactions on microscopic scales.
Finally, in section \ref{sec:Coasening}, we investigate the coarsening dynamics for both, wave and cluster lane patterns within the density segregated phase.

Section \ref{sec:Outlook}, then summarizes our numerical approach, discusses our main findings, and gives an outlook of potential applications of \SNAKE. In particular, this section gives a comprehensive and self-contained discussion of the most pertinent results established in the more technical sections \ref{sec:Discretization} and \ref{sec:Applications}, and can be read independently of these technical parts. Section \ref{sec:Outlook} has been written to provide the reader with a direct and non-technical accessibility to the core results of this work.

\section{\label{sec:Discretization}Discretization of the Boltzmann Equation}
Here and in the following, we focus on a kinetic description for two-dimensional systems of actively propelled particles in an over-damped environment. As a simplifying assumption, we consider particles moving at constant speed $v$ but variable orientation $\theta$, such that each particle's velocity vector $\vec v$ can be written $\vec v=v\,\hat{\vec e}_{\theta}$ ($\hat{\vec e}_{\theta}$: unit vector). For particles with nearly constant propelling forces, this is a reasonable approximation in the ``Aristotlean regime'' where the dynamics is governed by a balance between propelling and dissipative forces and inertial effects can be neglected. 
Moreover, for sufficiently dilute systems, binary particle interactions are believed to dominate over higher order interactions and the spatio-temporal dynamics of the one-particle distribution function $f(\vec x,\theta,t)$ can be captured by means of the Boltzmann equation~\footnote{Here and in the following, we tacitly assume molecular chaos to apply to all systems considered in this work. For a critical assessment of the assumption of molecular chaos in the Boltzmann Equation refer to~\cite{Thuroff_2013,Hanke_2013}.}.

\subsection{\label{sec:ContinuousBoltzmannEq}Boltzmann equation for active particles}
The most general form of the Boltzmann equation reads
\begin{subequations}
\label{eq:Boltzmann}
\begin{equation}
\label{eq:Boltzmann1}
\partial_t f(\vec x,\theta,t) + v\,\hat{\vec e}_{\theta}\cdot\nabla f(\vec x,\theta,t) = I_{\text{source}}\left[f;\,\theta\right].
\end{equation}
The density of particles in the phase space element $\bm\omega=[\vec x,\,\vec x + d\vec x]\times[\theta,\,\theta+d\theta]$ is being convected due the particles' active propulsion (left hand side), and is, simultaneously, subject to sudden changes due to particle collisions, chemical reactions, etc.\, which is captured on the right hand side of Eq.~\eqref{eq:Boltzmann1} in the form of source terms $I_{\text{source}}$. To make contact to previous works on the Boltzmann equation for active systems~\cite{Bertin_short,Bertin_long,Peshkov:2012uu}, we consider two separate contributions to the source term,
\begin{equation}
\label{eq:Boltzmann2}
I_{\text{source}}\left[f;\,\theta\right] = I_{\text{sd}}\left[f;\,\theta\right] + I_{\text{c}}\left[f;\,\theta\right],
\end{equation}
which take into account rotational diffusion due to a fluctuating background and / or noisy propelling mechanism, $I_{\text{sd}}$, and binary particle collisions, $I_{\text{c}}$, respectively.

To be specific, rotational diffusion is implemented using a constant ``tumbling'' rate $\lambda$, and drawing the corresponding angular shifts from a given probability distribution $p_0$. The self-diffusion integral $I_{\text{sd}}$ then takes the following form (omitting any explicit reference to space and time coordinates for brevity):
\begin{equation}
\label{eq:Isd}
\begin{split}
&I_{\text{sd}}\left[f;\,\theta\right] =\\
&-\lambda\,f(\theta) +  \lambda\int_{-\pi}^{\pi}[d\phi]_{(m)}\,f(\phi) \, p_0(\theta-\phi+2m\pi),
\end{split}
\end{equation}
where we introduced the short hand notation
\begin{equation*}
\int_{-\pi}^{\pi}[d\phi]_{(m)}\equiv\sum_{m=-\infty}^{\infty} \int_{-\pi}^{\pi}\,d\phi.
\end{equation*}
to take into account the $2\pi$-periodicity of $f(\theta)$.
In Eq.~\eqref{eq:Isd}, the two terms on the right hand side quantify the diffusion-related loss and gain of particles in the phase space element $\bm\omega$, respectively.

To asses the effect of binary particle collisions, we assume that the one-particle distribution function $f(\vec x,\theta,t)$ varies on time and length scales which are large compared to the typical time between particle collisions and typical inter-particle separations, respectively. Therefore, on the scale of the Boltzmann equation~\eqref{eq:Boltzmann1}, the impact of collisions on the distribution of angles $\theta$ can be formulated locally in space. To capture the effect of binary collisions, we resort to a collision rule, i.e.\ a model-specific mapping $\bm\theta_w$ between pre- and post-collisional angles: 
\begin{equation*}
\Bigl(\theta_1,\theta_2\Bigr)\mapsto \Bigl( \theta_w^{(1)}(\theta_1,\theta_2),\theta_w^{(2)}(\theta_1,\theta_2) \Bigr).
\end{equation*}
Further, we assume that the collision process is subject to noise, which causes a rotation of the post-collisional angles $\theta_w^{(1/2)}$ by some random amount drawn from a probability distribution $p$. We then arrive at the following form of the collision integrals $I_{\text{c}}$ (again, omitting any explicit reference to space and time coordinates):
\begin{equation}
\label{eq:Ic}
\begin{split}
&I_{\text{c}}\left[f;\,\theta\right] =\\
& -f(\theta)\int_{\theta-\psi}^{\theta+\psi}d\phi\,\Gamma(\theta-\phi)f(\phi)\\
&+\int_{-\pi}^{\pi}[d\phi_1]_{(m)}\int_{\phi_1-\psi}^{\phi_1+\psi}d\phi_2\,\Gamma(|\phi_2-\phi_1|)\,f(\phi_1)f(\phi_2)\\
&\times\frac{1}{2}\left[p\left(\theta-\theta_w^{(1)}+2m\pi\right) + p\left(\theta-\theta_w^{(2)}+2m\pi\right)\right],
\end{split}
\end{equation}
\end{subequations}
where $\Gamma(|\phi_1-\phi_2|)$ is a phase space factor (``Boltzmann Sto{\ss}zylinder''), quantifying the rate of collisions between particles with orientations $\phi_1$ and $\phi_2$. The first term on the right hand side of Eq.~\eqref{eq:Ic} gives the rate at which the phase space element $\bm\omega$ loses particles due to binary collisions. Similarly, the second term quantifies the rate at which particles are scattered into $\bm\omega$. For the sake of greater generality, we further introduced the parameter $\psi\in[0,\pi]$ in Eq.~\eqref{eq:Ic}, to account for systems with limited angular interaction ranges \cite{Thuroff_2013}.

Eqs.~\eqref{eq:Boltzmann} have been investigated in a number of previous studies~\cite{Bertin_short,Bertin_long,Peshkov:2012uu,Peshkov:2012tu,Weber_NJP_2013,Thuroff_2013} by mapping the mesoscopic Boltzmann equation~\eqref{eq:Boltzmann1} to a set of hydrodynamic equations of motion for the first few moments of the one-particle distribution function $f(\vec x,\theta,t)$. As discussed in the introduction, this mapping is based on a truncation of terms in Fourier space representation of the Boltzmann equation which itself rests on a perturbation ansatz for the Fourier modes around an isotropic base state. To which extent such an approach yields reliable results in the context of ordered systems remains an open question; cf. Ref.~\cite{Ihle:2013jg}. In the following sections, we take a different route and focus on a kinetic description for active systems by a direct numerical solution of the Boltzmann equation~\eqref{eq:Boltzmann}.

Before we derive a discretized version of Eqs.~\eqref{eq:Boltzmann}, one last remark is in order: Since Eqs.~\eqref{eq:Boltzmann2}--\eqref{eq:Ic} imply $\int_{-\pi}^{\pi}d\theta\,I_{\text{source}}[f;\,\theta]=0$, Eq.~\eqref{eq:Boltzmann1} gives rise to a conservation law for the local particle density $\rho(\vec x,t)=\int_{-\pi}^{\pi}d\theta\,f(\vec x,\theta,t)$:
\begin{equation}
\label{eq:LocalParticleConservation}
\partial_t\rho(\vec x,t)+\nabla\cdot\vec g(\vec x,t)=0,
\end{equation}
where 
\begin{equation}
\label{eq:def_momentum}
\vec g(\vec x,t)=v\int_{-\pi}^{\pi}d\theta\,\hat{\vec e}_{\theta}\,f(\vec x,\theta,t)
\end{equation}
denotes the local momentum density. In what follows, we will restrict ourselves to the discussion of closed systems in which the particle flux across its boundaries vanishes \footnote{Extension of the \SNAKE~algorithm to situation of a non-vanishing flux at the boundaries is straightforward}. Eq.~\eqref{eq:LocalParticleConservation} then entails conservation of the overall particle density 
\begin{equation}
\label{eq:ParticleConservation}
\rho_0 = \frac{1}{||\Omega||}\, \int_{\Omega}d^2x\,\rho(\vec x,t)=\text{const},
\end{equation}
where $\Omega$ denotes the system's two-dimensional spatial domain, and $||\Omega||$ its volume.

\subsection{\label{sec:DimlessBE}Dimensionless Boltzmann Equation}
We start by rewriting Eqs.~\eqref{eq:Boltzmann} in dimensionless form, in order to reduce the number of parameters. To this end, we measure time in units of the self-diffusion time scale $\lambda^{-1}$, distances in units of the ``ballistic flight length'' $v\,\lambda^{-1}$, and the one-particle distribution function $f$ in units of the (constant) overall particle density $\rho_0$ [cf. Eq.~\eqref{eq:ParticleConservation}]. Finally, the phase space factor $\Gamma$ quantifies the rate particle collisions and is, therefore, proportional to the particle speed $v$ and the particle extension $d$. Dimensional analysis then dictates $\tilde\Gamma=d\,v$. We thus arrive at the following convention of units
\begin{subequations}
\label{eq:Rescalings}
\begin{eqnarray}
t &\rightarrow& t \cdot \lambda^{-1},\\
\vec x &\rightarrow& \vec x \cdot v\,\lambda^{-1},\\
\label{eq:Rescalings3}
f &\rightarrow& f \cdot \rho_0,\\
\Gamma &\rightarrow& \Gamma \cdot d\,v,\\
\rho &\rightarrow& \rho\cdot\lambda\,(d\,v)^{-1},
\end{eqnarray}
\end{subequations}
which will be used throughout this work. The Boltzmann equation \eqref{eq:Boltzmann1} then takes the dimensionless form
\begin{equation}
\label{eq:Boltzmann_rescaled}
\partial_t f(\vec x,\theta,t) + \hat{\vec e}_{\theta}\cdot\nabla f(\vec x,\theta,t) = I_{\text{sd}}\left[f;\,\theta\right] + \rho_0\,I_{\text{c}}\left[f;\,\theta\right].
\end{equation}
Here the expressions for $I_{\text{sd}}$ and $I_{\text{c}}$ are given in Eqs.~\eqref{eq:Isd} (with $\lambda=1$) and~\eqref{eq:Ic}, respectively, with $f$ and $\Gamma$ in their dimensionless forms.

In the present system of units, the dimensionless density parameter $\rho_0$ 
measures the rate of collisions [$\rho_0\,d\,v$] relative to the rate of self-diffusion [$\lambda$]. As an aside we note that in a regime of high densities and / or large ballistic flight lengths with $\rho_0\gg1$, the characteristic scales introduced above are no longer meaningful. In such cases, the characteristic time scale would be set by the typical time between subsequent collisions, $(\rho_0\,d\,v)^{-1}$, the characteristic length scale by the typical inter-particle distance $(\rho_0\,d)^{-1}$. In this limit, self diffusion can be neglected and the problem becomes independent of the overall particle density.

\subsection{Discretization scheme}
Having established a dimensionless form for the Boltzmann equation for active, two-dimensional systems, Eq.~\eqref{eq:Boltzmann_rescaled}, this section is devoted to the derivation of a discretized version of Eq.~\eqref{eq:Boltzmann_rescaled}.
In order to keep the notation compact, we use ``co- and contravariant indices'' such that summation over repeated co- and contravariant index pairs is implied. Throughout this work, we reserve greek letters for spatial indices, and latin letters for angular indices. Moreover, we use the symbol $||\dots||$ to denote the measure of spatial domains and angular intervals.

\begin{figure}[t]
\begin{center}
\includegraphics{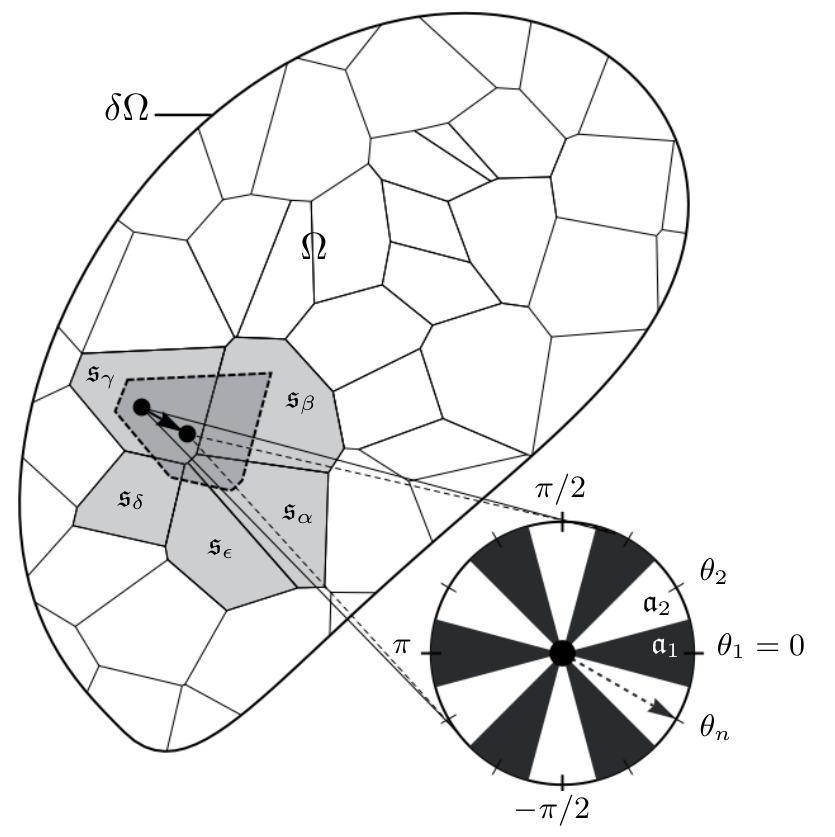}
\caption{
\emph{Spatial and angular tessellations:}
The system's spatial domain $\Omega$ is discretized using an arbitrary tessellation $\{\mathfrak{s}^{\alpha}\}_{\alpha}$ obeying Eqs.~\eqref{eq:SpatialTessellation}. A set of angular channels according to the tessellation~\eqref{eq:SpecificAngularTessellation} is attached to each spatial cell $\mathfrak{s}^{\gamma}$.
\emph{Spatial transport:}
Convection in the bulk of the system along the direction $\theta_n$ (indicated by the dashed arrow) amounts to translating each spatial cell $\mathfrak{s}^{\gamma}$ by the convection length $\tau$. The \emph{convective transport operator} $\mathcal{T}_{n \gamma}^{\alpha k}$, Eq.~\eqref{eq:def_conv_matrix}, then redistributes the current value of $f_n^{\gamma}$ among all neighboring cells according to their intersection with the convected cell. 
}
\label{fig:tessellations}
\end{center}
\end{figure}

To discretize the spatial domain $\Omega$, we consider a general tessellation by means of an arbitrary set of connected domains $\{\mathfrak s^{\alpha}\}_{\alpha=1,\dots,M}$ such that they are a partition of $\Omega$, i.e. [$\delta_{\beta}^{\alpha}$: Kronecker delta]
\begin{subequations}
\label{eq:SpatialTessellation}
\begin{eqnarray}
\mathfrak s^{\alpha} &\subset& \Omega,\\
||\mathfrak s^{\alpha} \cap \mathfrak s^{\beta}|| &=&||\mathfrak s^{\beta}||\,\delta_{\beta}^{\alpha},\\
\bigcup_{\alpha=1}^M\mathfrak s^{\alpha} &=& \Omega,
\end{eqnarray}
\end{subequations}
for all $\alpha,\beta\in\{1\dots M\}$. Similarly, to partition the angular range $A=(-\pi,\pi]$, we consider tessellations by means of a set of simply connected intervals $\{\mathfrak a_{n}\}_{n=1,\dots,K}$, such that
\begin{subequations}
\label{eq:AngularTessellation}
\begin{eqnarray}
\mathfrak a_{n} &\subset& A,\\
||\mathfrak a_{n} \cap \mathfrak a_{m}|| &=&||\mathfrak a_{m}||\,\delta_{n}^{m},\\
\bigcup_{n=1}^K\mathfrak a_{n} &=& A,
\end{eqnarray}
\end{subequations}
for all $n,m\in\{1,\dots, K\}$; cf Fig.~\ref{fig:tessellations}. The particular choices for the spatial tessellations are of no importance for the general development of our discretization scheme. However, some care must be taken in the construction of the angular tessellations. To avoid creating an artificial angular bias, the angles $\{\theta_n\}_{n=1,\dots,K}$, used to represent the partition $\{\mathfrak a_{n}\}_{n=1,\dots,K}$, should sum up to zero. In addition, the angular partition should be constructed such that any special lattice found in the spatial tessellations (e.g. when space-filling, regular partitions are being used) should be reflected in any particular choice of angular tessellations. In what follows, we will use
\begin{equation}
\label{eq:SpecificAngularTessellation}
\mathfrak a_{n} = \left(\theta_n-\frac{\pi}{K},\,\theta_n+\frac{\pi}{K}\right], \qquad\theta_n=\frac{2\pi n}{K},
\end{equation}
and choose $K\in4\mathbb N$ such that (i) the angular partition $\{\mathfrak a_{n}\}_{n}\supset\{0,\pi,\pm\pi/2\}$ and (ii) for each angular domain $\mathfrak a_{i}\in\{\mathfrak a_{n}\}_{n}$ there exists a domain $\mathfrak{a}_{j}\in\{\mathfrak{a}_{n}\}_{n}$ such that $\theta_i=\theta_j+\pi$ (modulo $2\pi$). This then implies 
\begin{equation}
||\mathfrak a_{n}|| = \frac{2\pi}{K} \equiv \epsilon_{\theta}.
\end{equation}

We proceed by defining indicator functions for the spatial partition of $\Omega$
\begin{equation}
\label{eq:SpatialFilter}
\mathcal{F}_{\alpha}(\vec x) \equiv
\begin{cases}
1, & \text{if } \vec x \in \mathfrak s^{\alpha},\\
0, & \text{else},
\end{cases}
\end{equation}
and the angular partition of $(-\pi,\pi]$,
\begin{equation}
\label{eq:AngularFilter}
\mathcal{F}^n(\theta) \equiv
\begin{cases}
1, & \text{if } \theta \in \mathfrak a_{n},\\
0, & \text{else},
\end{cases}
\end{equation}
which can be used to write
\begin{equation}
\label{eq:fstepfct}
f(\vec x,\theta,t) \equiv f_k^{\alpha}(t)\,\mathcal{F}_{\alpha}(\vec x)\mathcal{F}^{k}(\theta).
\end{equation}
Here, the time dependent \emph{density matrix} $f_k^{\alpha}(t)\geq0$ constitutes a discrete representation of the continuous one-particle distribution function $f(\vec x,\theta,t)$. To assess the temporal evolution of the \emph{density matrix} $f_k^{\alpha}$, it suffices to establish a transformation law of this matrix under small time increments. This transformation law then provides us with an ``update rule'' to compute the temporal evolution $f_k^{\alpha}(t)$ iteratively, starting from some initial distribution $f_k^{\alpha}(0)$, and hence to numerically solve the Boltzmann equation \eqref{eq:Boltzmann_rescaled}. In the following, we restrict ourselves to considering the time evolution in the bulk of the system, postponing a detailed discussion of boundary conditions to section \ref{sec:BCs}.

\subsubsection{\label{sec:ConvectiveOperator}Convective transport operator}
Consider a small translation in the (rescaled) time coordinate, $t\rightarrow t+\tau$, where $\tau\ll1$. To assess the convective contribution to the time transformation law for the \emph{density matrix} $f_k^{\alpha}$, we replace the convective derivative on the left hand side of the Boltzmann equation~\eqref{eq:Boltzmann_rescaled} by finite differences. Using a forward difference to substitute for the partial time derivative, we get
\begin{subequations}
\label{eq:ConvectiveDerivative}
\begin{equation}
\label{eq:pdt}
\partial_tf(\vec x,\theta,t) \rightarrow \frac{f(\vec x,\theta,t+\tau)-f(\vec x,\theta,t)}{\tau}.
\end{equation}
Further, noting that $\hat{\vec e}_{\theta}\cdot\nabla f$ is the directional derivative of $f$ along $\hat{\vec e}_{\theta}$, we write
\begin{equation}
\label{eq:pdtconv}
\hat{\vec e}_{\theta}\cdot\nabla f(\vec x,\theta,t) \rightarrow\,\frac{f(\vec x,\theta,t)-f(\vec x-\tau\,\hat{\vec e}_{\theta},\theta,t)}{\tau},
\end{equation}
\end{subequations}
where now a backward difference is used to replace the spatial derivative in order to achieve cancellation of the terms $\propto f(\vec x,\theta,t)$. We chose equal step lengths in Eqs.~\eqref{eq:pdt} and~\eqref{eq:pdtconv} to account for the fact that the particles move at unit speed $v=1$ according to our units convention. To derive the time transformation law for the \emph{density matrix} $f_k^{\alpha}$ due to particle convection, we add Eqs.~\eqref{eq:pdt} and~\eqref{eq:pdtconv}, substitute the discretization ansatz~\eqref{eq:fstepfct} for the one-particle distribution function $f(\vec x,\theta,t)$, and project the resulting expression onto the discretized phase space volume $\bm\omega_n^{\alpha}=\mathfrak s^{\alpha}\otimes\mathfrak a_{n}$. Introducing the projection operator
\begin{equation}
\Pi_n^{\alpha}\equiv\frac{1}{\epsilon_{\theta}\,||\mathfrak{s}^{\alpha}||}\int_{\mathfrak s^{\alpha}}d^2x\int_{\mathfrak a_{n}}d\theta\equiv\int d\bm\tilde\omega_n^{\alpha},
\end{equation}
we get
\begin{subequations}
\label{eq:convection}
\begin{equation}
\label{eq:convection1}
\begin{split}
&\Pi_n^{\alpha}\left\{\Bigl[\partial_t+\hat{\vec e}_{\theta}\cdot\nabla\Bigr] f(\vec x,\theta,t)\right\}\equiv\\
&\int d\bm\tilde\omega_n^{\alpha}\,\mathcal{F}^k(\theta)\Bigl[f_k^{\gamma}(t+\tau)\mathcal{F}_{\gamma}(\vec x)-f_k^{\gamma}(t)\mathcal{F}_{\gamma}(\vec x-\tau\,\hat{\vec e}_{\theta})\Bigr]/\tau\\
&=\Bigl[f_n^{\alpha}(t+\tau)-\mathcal{T}_{n\gamma}^{\alpha k}\,f_k^{\gamma}(t)\Bigr]/\tau,
\end{split}
\end{equation}
where we introduced the \emph{convective transport operator}
\begin{equation}
\label{eq:def_conv_matrix}
\mathcal{T}_{n \gamma}^{\alpha k} = \delta_n^k\,\int d\bm\tilde\omega_n^{\alpha}\,\mathcal{F}_{\gamma}(\vec x-\tau\,\hat{\vec e}_{\theta}),
\end{equation}
\end{subequations}
which quantifies the time propagation of the \emph{density matrix} $f_n^{\alpha}$, due to the convective transport of particles. 
From Eq.~\eqref{eq:def_conv_matrix} we can extract a direct geometrical interpretation for the implementation of convective transport processes via $\mathcal{T}_{n \gamma}^{\alpha k}$; cf. Fig.~\ref{fig:tessellations}: The \emph{convective transport operator} is computed by ``convecting'' the  indicator function $\mathcal{F}_{\gamma}$ for the spatial domain $\mathfrak{s}_{\gamma}$ along $\theta_n$, and integrating the resulting function over the domain $\mathfrak{s}_{\alpha}$. Geometrically, this amounts to computing the area of intersection between $\mathfrak{s}_{\alpha}$ and a convected copy of $\mathfrak{s}_{\gamma}$ divided by the volume of the phase space element $\bm\omega_{n}^{\alpha}$. The operator element $\mathcal{T}_{n \gamma}^{\alpha k}$ thus measures the number particles being transported from $\bm\omega_n^{\gamma}$ to $\omega_{n}^{\alpha}$ during $(t-dt,t)$. More precisely, $\mathcal{T}_{n \gamma}^{\alpha k}$ gives the increase in particle density in the phase space element $\omega_{n}^{\alpha}$ relative to the particle density in the phase space element $\bm\omega_n^{\gamma}$ at time $t-dt$ due to convective coupling between these elements in the time interval $(t-dt,t)$.

A few remarks on the properties of the \emph{convective transport operator} are in order. First of all, since, in the absence of self-diffusion and collisions, the particles maintain their current orientation, the \emph{convective transport operator} is diagonal with respect to the angular indices for convective processes in the bulk of the system. This is, however, no longer true for convection in the vicinity of a (non-periodic) boundary, where particles generally reorient due to collisions with boundary walls. A corresponding redefinition of the \emph{convective transport operator} will be given in section \ref{sec:BCs}, where we discuss the implementation of non-periodic boundary conditions. Secondly, since $\mathcal{T}_{n\gamma}^{\alpha k}$ reduces to the identity operator in the limit $\tau\rightarrow0$, the \emph{convective transport operator} becomes ``quasi diagonal'' in the spatial indices for sufficiently small time steps $\tau$. More precisely, while the diagonal elements $\mathcal{T}_{n\alpha}^{\alpha n}$ are $\mathcal{O}(1)$, off-diagonal elements are $\mathcal{O}(\tau)$, i.e. convective coupling between neighboring elements becomes increasingly small as $\tau\rightarrow0$. This, in turn, has important consequences for the computation of the angular self-diffusion and collision contributions to the time evolution of the \emph{density matrix} $f_n^{\alpha}$, which themselves are $\mathcal{O}(\tau)$ as will be shown below. Thus, up to linear order in $\tau$, convective processes decouple from angular diffusion and collision, such that the latter processes can be computed independently inside each spatial domain $\mathfrak{s}_{\alpha}$.

\subsubsection{\label{sec:sd_operator}Self diffusion operator}
To discretize the self-diffusion integral $I_{\text{sd}}$, we proceed along the same lines as in the preceding section and substitute the discrete representation of $f(\vec x,\theta,t)$, Eq.~\eqref{eq:fstepfct}, and project the resulting expression onto the phase space volume $\bm\omega_n^{\alpha}$. Using the analytical form of the self-diffusion matrix, Eq.~\eqref{eq:Isd}, we find
\begin{subequations}
\begin{equation}
\label{eq:SDdiscretized}
\begin{split}
\Pi_{n}^{\alpha}\,I_{\text{sd}}&\equiv-f_n^{\alpha}(t)\,+\int d\bm\tilde\omega_n^{\alpha} \int_{-\pi}^{\pi}\,[d\phi]_{(m)}\\
&\times\mathcal{F}^{k}(\phi)\,f_k^{\gamma}(t)\,\mathcal{F}_{\gamma}(\vec x) \, p_0(\theta-\phi+2m\pi)\\[.2cm]
&=\mathcal{D}_n^k\,f_k^{\alpha}(t),
\end{split}
\end{equation}
where the matrix elements of the \emph{self-diffusion operator}
\begin{equation}
\label{eq:SDmatrix}
\mathcal{D}_n^k = \int_{\mathfrak a_{n}} \frac{d\theta}{\epsilon_{\theta}} \int_{\mathfrak a_{k}}\,[d\phi]_{(m)}\,p_0(\theta-\phi+2m\pi)-\delta_n^k
\end{equation}
\end{subequations}
quantify the net gain of particles (per unit time) in the angular element $\mathfrak a_n$, by diffusive scattering from the angular element $\mathfrak a_k$. 

\subsubsection{\label{sec:col_operator}Collision operator}
Finally, the discretized version of the collision integral $I_{\text{c}}$ defines the \emph{collision operator} $\mathcal{C}_n^{k l}$, which can be calculated using the standard procedure outlined in the previous two subsections. Starting loss term in Eq.~\eqref{eq:Ic}, we obtain:
\begin{subequations}
\label{eq:CollisionLoss}
\begin{equation}
\begin{split}
\Pi_{n}^{\alpha}\,I_{\text{c}}^{\text{(loss)}}&\equiv-\int d\bm\tilde\omega_n^{\alpha}\int_{\theta-\psi}^{\theta+\psi}d\phi\,\Gamma(|\theta-\phi|)\\
&\qquad\times\mathcal{F}_{\gamma}(\vec x)\mathcal{F}_{\delta}(\vec x)f_k^{\gamma}(t)f_l^{\delta}(t)\mathcal{F}^{k}(\theta)\mathcal{F}^{l}(\phi)\\[.2cm]
&=-f_n^{\alpha}(t)\,\mathcal{S}_n^l\,f_l^{\alpha}(t),
\end{split}
\end{equation}
where the matrix elements
\label{eq:SMatrix1}
\begin{equation}
\mathcal{S}_n^l=\int_{\mathfrak a_{n}}\frac{d\theta}{\epsilon_{\theta}}\int_{\theta-\psi}^{\theta+\psi}d\phi\,\Gamma(|\theta-\phi|)\mathcal{F}^{l}(\phi)
\end{equation}
\end{subequations}
measure the loss of particles (per unit time) in the angular element $\mathfrak a_n$ due to collisions with particles in the element $\mathfrak a_k$.

Similarly, considering the gain term of Eq.~\eqref{eq:Ic}, we find:
\begin{subequations}
\label{eq:CollisionGain}
\begin{equation}
\begin{split}        
&\Pi_{n}^{\alpha}\,I_{\text{c}}^{\text{(gain)}}\equiv\\
&\int d\bm\tilde\omega_n^{\alpha}\int_{-\pi}^{\pi}[d\phi_1]_{(m)}\int_{\phi_1-\psi}^{\phi_1+\psi}d\phi_2\,\Gamma(|\phi_2-\phi_1|)\\
&\times \mathcal{F}_{\gamma}(\vec x)\mathcal{F}_{\delta}(\vec x)f_k^{\gamma}(t)f_l^{\delta}(t)\mathcal{F}^{k}(\phi_1)\mathcal{F}^{l}(\phi_2)\\
&\times \frac{1}{2}\left[p\left(\theta-\theta_w^{(1)}+2\pi m\right)+p\left(\theta-\theta_w^{(2)}+2\pi m\right)\right]\\[.2cm]
&=f_k^{\alpha}(t)\mathcal{S}_n^{k l}f_l^{\alpha}(t),
\end{split}
\end{equation}
where the matrix elements
\begin{equation}
\label{eq:SMatrix2}
\begin{split}
&\mathcal{S}_n^{k l}=\\
&\int_{\mathfrak a_{n}} \frac{d\theta}{\epsilon_{\theta}}\int_{\mathfrak a^{k}}[d\phi_1]_{(m)}\int_{\phi_1-\psi}^{\phi_1+\psi}d\phi_2\,\Gamma(|\phi_2-\phi_1|)\mathcal{F}^{l}(\phi_2)\\
&\times \frac{1}{2}\left[p\left(\theta-\theta_w^{(1)}+2\pi m\right)+p\left(\theta-\theta_w^{(2)}+2\pi m\right)\right]
\end{split}
\end{equation}
\end{subequations}
count the number of particles (per unit time) being scattered into the angular element $\mathfrak a_n$ as a result of binary collisions between particles with angles $\phi_1\in\mathfrak a_k$ and $\phi_2\in\mathfrak a_l$.

Combining Eqs.~\eqref{eq:CollisionLoss} and \eqref{eq:CollisionGain}, we find
\begin{subequations}
\label{eq:CollisionTerms}
\begin{equation}
\Pi_{n}^{\alpha}\,I_{\text{c}}\equiv f_k^{\alpha}(t)\mathcal{C}_n^{k l}f_l^{\alpha}(t),
\end{equation}
with the \emph{collision operator} $\mathcal{C}_n^{k l}$ given by:
\begin{equation}
\label{eq:CollisionOperator}
\mathcal{C}_n^{k l}=\mathcal{S}_n^{k l}-\mathcal{S}_n^l\,\delta_n^k.
\end{equation}
\end{subequations}

\subsubsection{\label{sec:UpdateRule}Discrete time transformation law and summation rules}
Having discretized the various terms in the Boltzmann equation, Eq.~\eqref{eq:Boltzmann_rescaled}, we can now combine our findings from the previous three subsections, to arrive at the following transformation law, capturing the evolution of the \emph{density matrix} $f_n^{\alpha}(t)$ under a small transformation in time, $t\rightarrow t+\tau$,
\begin{equation}
\label{eq:UpdateRule}
f_n^{\alpha}(t+\tau)=\Bigl[\mathcal{T}_{n\gamma}^{\alpha k} +  \tau\Bigl(\mathcal{D}_n^k+\rho_0\,\mathcal{C}_n^{k l}f_l^{\alpha}(t)\Bigr)\delta_{\gamma}^{\alpha}\Bigr]\,f_k^{\gamma}(t),
\end{equation}
with the various operators, acting on the \emph{density matrix} $f_n^{\alpha}(t)$, given in Eqs.~\eqref{eq:def_conv_matrix}, \eqref{eq:SDmatrix}, and \eqref{eq:CollisionOperator}. Eq.~\eqref{eq:UpdateRule} defines a recursive relation, which allows us to numerically compute the solutions of the Boltzmann equation, Eq.~\eqref{eq:Boltzmann_rescaled}, given initial conditions $f_n^{\alpha}(0)$. It, therefore, constitutes a discrete version of the Boltzmann equation for active systems, as introduced in section~\ref{sec:ContinuousBoltzmannEq}.

We conclude the current section by demonstrating that the particle conservation property, inherent in the original Boltzmann equation , Eq.~\eqref{eq:Boltzmann_rescaled}, is being preserved by the transformation law, Eq.~\eqref{eq:UpdateRule}. To this end, we use the following operator summation rules, which themselves follow directly from the respective definitions of the corresponding operators:
\begin{subequations}
\label{eq:SummationRules}
\begin{eqnarray}
\label{eq:SummationRules1}
||\mathfrak s_{\alpha}||\,||\mathfrak a^n||\mathcal{T}_{n\gamma}^{\alpha k}&=&||\mathfrak s_{\gamma}||\,||\mathfrak a^k||,\\
\label{eq:SummationRules2}
||\mathfrak a^n||\,\mathcal{D}_{n}^{k}&=&0,\\
\label{eq:SummationRules3}
||\mathfrak a^n||\,\mathcal{C}_{n}^{k l}&=&0,
\end{eqnarray}
\end{subequations}
In particular, Eq.~\eqref{eq:SummationRules1} expresses the fact that convective streaming in the bulk conserves the volume of the phase space elements $\bm\omega_{\gamma}^k=\mathfrak s_{\gamma}\otimes\mathfrak a^k$. 
Eqs.~\eqref{eq:SummationRules2} and~\eqref{eq:SummationRules3} capture the particle conserving properties of the \emph{self diffusion} and \emph{collision operators}, by asserting that the total gain of particles in each angular element is balanced by the total loss of particles in all remaining angular elements, and vice versa. Now, using Eqs.~\eqref{eq:SummationRules}, we find
\begin{equation}
\begin{split}
&N(t+\tau)=||\mathfrak s_{\alpha}||\,||\mathfrak a^{n}||\,f_n^{\alpha}(t+\tau)\\
&=||\mathfrak s_{\alpha}||\,||\mathfrak a^{n}||\,\Bigl[\mathcal{T}_{n\gamma}^{\alpha k} +  \tau\Bigl(\mathcal{D}_n^k+\rho_0\,\mathcal{C}_n^{k l}f_l^{\alpha}(t)\Bigr)\delta_{\gamma}^{\alpha}\Bigr]\,f_k^{\gamma}(t)\\
&=||\mathfrak s_{\gamma}||\,||\mathfrak a^{k}||\,f_k^{\gamma}(t)=N(t),
\end{split}
\end{equation}
i.e.\ conservation of the total number of particles. 

Finally, we note that, without further simplification, the computation time for an algorithm based on Eq.~\eqref{eq:UpdateRule} is of order $\mathcal{O}(K^3)$ \footnote{The rate limiting step here is the implementation of particle collisions. It takes $\mathcal{O}(K^2)$ operations to update each one of $K$ angular channels of the local \emph{density matrix} $f_n^{\alpha}$.}, and is thus effectively set by the number of angular slices $K$. 
 In the absence of collision noise, on the other hand, the transition probabilities in the definition of the \emph{collision operator} [cf. Eq.~\eqref{eq:SMatrix2}] reduce to delta peaks about $\theta_w^{(1/2)}$, rendering the implementation of Eq.~\eqref{eq:UpdateRule} of order $\mathcal{O}(K^2)$.
Unless stated otherwise, we will use $K=32$ in what follows.
To test whether this is an appropriate choice for the angular resolution, we computed the spatially homogeneous solution of the \emph{density matrix} $f_n^{\alpha}$ for a range of values $K\in\{32,\dots,128\}$ and a number of different density levels, using a $1\times1$ spatial grid. In fact, we observe no significant quantitative improvement upon increasing the number of angular slices in the presence of collision noise; cf. Fig.~\ref{fig:phase_diagram_cuts}(a).
The stationary, spatially homogeneous solution for the one-particle distribution function $f(\theta)$ as depicted in Fig.~\ref{fig:phase_diagram_cuts}(a) has previously been observed
numerically~\cite{Aronson_MT} and predicted analytically~\cite{Ben-Naim_2006}.

\subsection{\label{sec:BCs}Boundary conditions}

The analytical form of Eq.~\eqref{eq:UpdateRule} applies equally well in the bulk and at the boundaries of the system. For non-periodic boundaries, however, the \emph{convective transport operator} $\mathcal{T}_{n\gamma}^{\alpha k}$ differs from its bulk form, Eq.~\eqref{eq:def_conv_matrix}, and has to be modified. In this section, we start by discussing these modifications in abstract terms, and will later apply the generalized form of the \emph{convective transport operator} to the specific case of reflective boundaries.

In the following, we shall confine ourselves to the discussion of fixed boundaries with a model specific reflection rule. Specifically, we assume particles to undergo a spontaneous reorientation upon collisions with the boundary,
\begin{equation}
\theta \rightarrow \theta_{\text{BC}}(\theta),
\end{equation}
where $\theta$ and $\theta_{\text{BC}}(\theta)$ denote the particle's orientation before and after the boundary contact, and where $\theta_{\text{BC}}$ can be either a deterministic or a random variable. For the sake of clarity, we will discuss deterministic particle-boundary interactions. The treatment of stochastic boundary conditions follows along similar lines.

To the considered order, self-diffusion processes and particle-particle collisions are evaluated at a fixed point in space and time in Eq.~\eqref{eq:UpdateRule}. To account for the presence of boundaries, we are, therefore, led to reconsider the derivation of the \emph{convective transport operator} $\mathcal{T}_{n\gamma}^{\alpha k}$; cf. section \ref{sec:ConvectiveOperator}. For small but finite values of the time shift $\tau$, we have to redefine the convective derivative, Eqs.~\eqref{eq:ConvectiveDerivative}, for all
\begin{equation}
\vec x\in\mathbb B_{\tau}=\left\{x\,;\,\min_{\vec x'\in\,\delta\Omega}|\vec x-\vec x'|<\tau\right\}.
\end{equation}
Here $\mathbb B_{\tau}$ denotes the set of all points in space which are within ``convective reach'' of the boundary $\delta\Omega$ of the system's domain $\Omega$ within the next time interval $[t,t+\tau]$. An appropriate redefinition of the \emph{convective transport operator} can then be given using \emph{``inverse propagation functions''}:
\begin{subequations}
\label{eq:inverse_propagation_functions}
\begin{eqnarray}
\nonumber
\vec c_{s}^{-1}(\vec x,\theta; \tau) &\equiv& \text{particle position at time $t$, given that the}\\
\nonumber
&& \text{particle is located at $\vec x$ and oriented}\\
&& \text{along $\hat{\vec e}_{\theta}$ at time $t+\tau$,}\\
\nonumber
c_{a}^{-1}(\vec x,\theta; \tau) &\equiv& \text{particle orientation at time $t$, given that}\\
\nonumber
&& \text{the particle is located at $\vec x$ and oriented}\\
&& \text{along $\hat{\vec e}_{\theta}$ at time $t+\tau$.}
\end{eqnarray}
\end{subequations}
Using these \emph{inverse propagation functions}, we redefine the finite difference approximation of the convective derivative~\eqref{eq:ConvectiveDerivative} as follows:
\begin{equation}
\begin{split}
&[\partial_t+\hat{\vec e}_{\theta}\cdot\nabla] \, f(\vec x,\theta,t) \rightarrow\\[.2cm]
& \frac{f(\vec x,\theta,t+\tau)-f(\vec c_{s}^{-1}(\vec x,\theta; \tau),c_{a}^{-1}(\vec x,\theta; \tau),t)}{\tau},
\end{split}
\end{equation}
which reduces to the bulk form, Eq.~\eqref{eq:ConvectiveDerivative}, upon substituting the inverse bulk propagation functions $\vec c_{s}^{-1}(\vec x,\theta; \tau)=\vec x-\tau\hat{\vec e}_{\theta}$, and $c_{a}^{-1}(\vec x,\theta; \tau) = \theta$. Using this generalized finite difference representation to replace the convective derivative, we can proceed along the same lines of section \ref{sec:ConvectiveOperator} to arrive at the following, generalized form of the \emph{convective transport operator}
\begin{equation}
\label{eq:def_conv_matrix_generic}
\mathcal{T}_{n \gamma}^{\alpha k} = \int\frac{d\bm\tilde\omega_n^{\alpha}}{\epsilon_{\theta}\,||\mathfrak s^{\alpha}||} \,\mathcal{F}_{\gamma}(\vec c_{s}^{-1}(\vec x,\theta; \tau))\mathcal{F}^k(c_{a}^{-1}(\vec x,\theta; \tau)).
\end{equation}
Since $c_{a}^{-1}=\theta_{\text{BC}}^{-1}\neq\mathbb 1$ for particles undergoing a collision with the system's boundary during $[t,t+\tau]$, $\mathcal{T}_{n \gamma}^{\alpha k}$ is no longer diagonal in the angular indices for $\vec x \in \mathbb B_{\tau}$.

The generalized \emph{convective transport operator}, Eq.~\eqref{eq:def_conv_matrix_generic} lends itself to implement virtually arbitrary boundary conditions, which enter the Eq.~\eqref{eq:def_conv_matrix_generic} by an appropriate definition of the \emph{inverse propagation functions} \eqref{eq:inverse_propagation_functions}. To illustrate the general concept, we conclude this section with a brief discussion of the implementation of reflective boundary conditions.

After the collision of a particle with a reflective boundary, the particle's orientation changes according to
\begin{subequations}
\begin{equation}
\theta' = 2\,\phi_b(\vec x)-\theta,
\end{equation}
where $\phi_b(\vec x)$ denotes the local orientation of the system's boundary, and where $\theta'$ and $\theta$ denote the particle's orientation before and after the collision with the reflective boundary, respectively. Moreover, if a particle's state is described by $(\vec x,\,\theta)$ at time $t$ and is known to have undergone a collision with the system's boundary in the time interval $(t-\tau,\,t]$, its spatial position $\vec x'$ at time $t-\tau$ is given by
\begin{equation}
\vec x' = R_{\phi_b}\,(\vec x-\tau\hat{\vec e}_{\theta}),
\end{equation}
\end{subequations}
where $R_{\phi_b}$ denotes a reflection with respect to the boundary's local tangent of orientation $\phi_b$. In summary, the \emph{inverse propagation functions} in the case of reflective boundary conditions read:
\begin{subequations}
\label{eq:inverse_propagation_reflective_BC}
\begin{equation}
\vec c_{s}^{-1}(\vec x,\theta; \tau) = 
\begin{cases}
R_{\phi_b}\,(\vec x-\tau\hat{\vec e}_{\theta}), & \overline{\vec x-\tau\hat{\vec e}_{\theta},\vec x}\cap\delta\Omega\neq\varnothing,\\
\vec x-\tau\hat{\vec e}_{\theta}, & \text{else,}
\end{cases}
\end{equation}
and
\begin{equation}
c_{a}^{-1}(\vec x,\theta; \tau) = 
\begin{cases}
2\,\phi_b(\vec x)-\theta, & \overline{\vec x-\tau\hat{\vec e}_{\theta},\vec x}\cap\delta\Omega\neq\varnothing,\\
\theta, & \text{else,}
\end{cases}
\end{equation}
\end{subequations}
where $\overline{\vec a,\vec b}$ denotes the line segment joining the points $\vec a$ and $\vec b$.

As a final remark we mention that Eqs.~\eqref{eq:inverse_propagation_reflective_BC} can be implemented straightforwardly using arbitrary boundary geometries. Although a detailed study on the impact of confining geometries on the behavior of active systems lies beyond the scope of our present work, \SNAKE~lends itself as a convenient starting point to address this important question. For sample simulations of active polar systems in circular and ring-like confinements, we refer the interested reader to the Supplemental Material \cite{Supplement}.

\section{\label{sec:Applications}Polar Systems: Phases and Transitions}
Having established a general framework to discretize the Boltzmann equation for active particles, Eq.~\eqref{eq:Boltzmann_rescaled}, we now discuss actual applications of \SNAKE~in the context of active polar systems. We will focus our considerations on a simple model system with a polar collision rule
\begin{equation}
\label{eq:half_angle_col_rule}
\Bigl(\theta_1,\theta_2\Bigr)\mapsto \left(\theta_w^{(1)} = \frac{\theta_1+\theta_2}{2},\,\theta_w^{(2)} =  \frac{\theta_1+\theta_2}{2} \right),
\end{equation}
and full angular interaction range. For the self-diffusion and collision noise probability functions $p$ and $p_0$, we choose zero-mean Gaussian probability distributions with variances $\sigma_0^2$ and $\sigma^2$, respectively.

This model has previously been proposed and studied by Bertin et al. \cite{Bertin_short,Bertin_long} using mainly analytical methods. Our purpose here is twofold. First, we can resort to a number of well established analytical results \cite{Bertin_long} for this model, as well as a number of previous studies of closely related systems \cite{Chate_2004,Chate_long,Ihle_2011,Ihle:2013jg}, to test for the accuracy and reliability of our computational scheme. Secondly, we can use \SNAKE~to go beyond the hydrodynamic picture of the onset of collective motion, as discussed in Refs.~\cite{Bertin_short,Bertin_long}, and study the nature of the various phases, as well as the ensuing transitions between them.

In the following sections, we will use the system of units introduced section \ref{sec:DimlessBE}; cf. Eqs.~\eqref{eq:Rescalings}. Moreover, we will use spatial tessellations in the form of regular two-dimensional arrays of $L\times L$ rectangular grid sites with linear extensions $|\mathfrak{s}_{\alpha}|=\text{const}\equiv\epsilon_x$, and periodic boundary conditions. Some care must be taken for the appropriate choice of the size of one ``Boltzmann element'', $\epsilon_x$: 
The spatial patterns typically formed by these systems at the onset of polar order, are known to develop rather steep wave fronts, with density profiles varying on length scales comparable to the collision length (i.e. the typical length a particle travels between successive collisions) \cite{Chate_long,Ihle:2013jg}. To be able to resolve such patterns with sufficient accuracy, we need $\epsilon_x\sim\rho_0^{-1}$. 
In all subsequent sections, we will use densities $\rho_0=\mathcal{O}(0.1-1)$, such that $\epsilon_x=5$ is an appropriate choice. Unless stated otherwise, $\epsilon_x=5$ and $L=100$ will be used throughout this work.

\subsection{\label{sec:PhaseDiagram}Phase diagram of stationary states}

\setlength{\tabcolsep}{-2.5em}
\begin{figure*}
\begin{tabular}{ccc}
\includegraphics{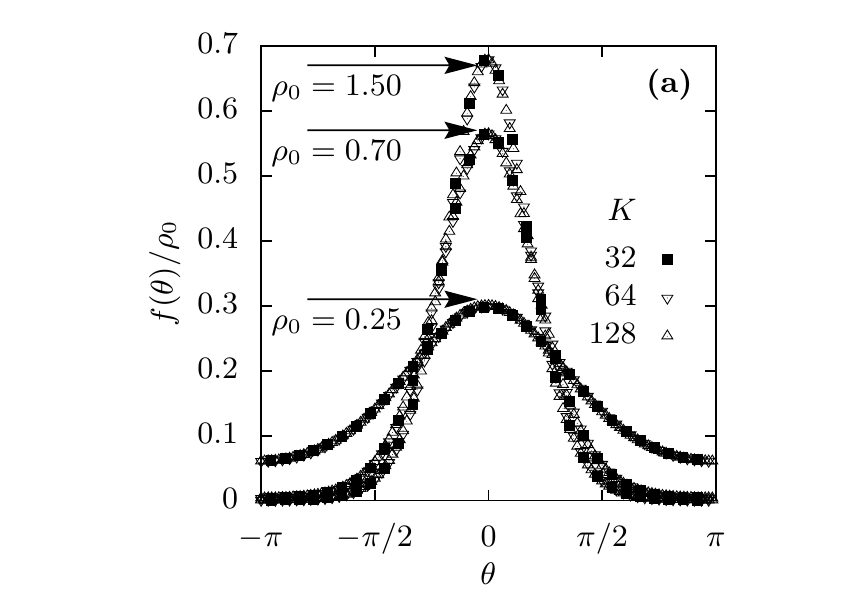} & \includegraphics{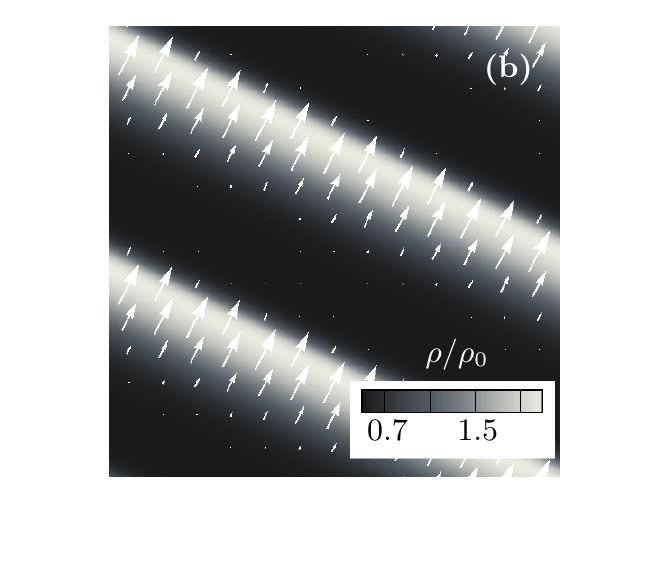} & \includegraphics{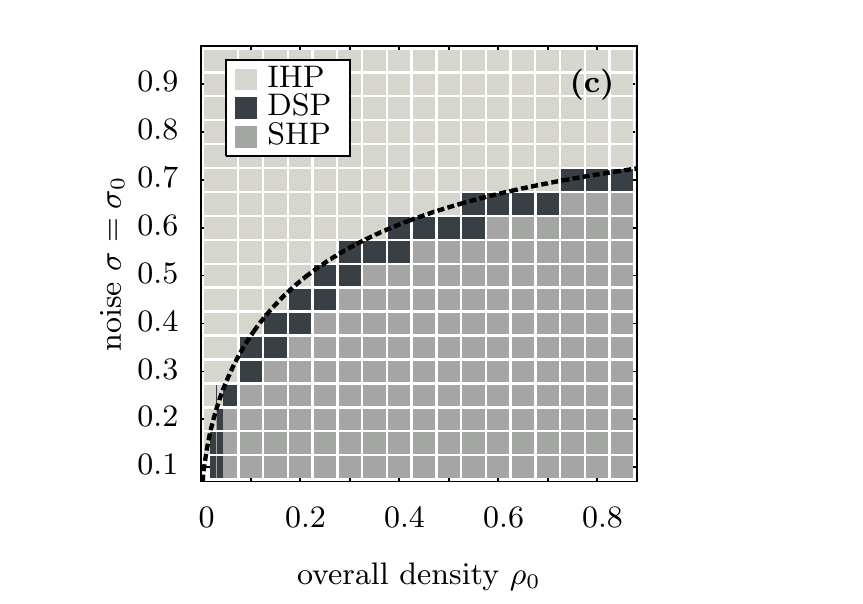}
\end{tabular}
\caption{
\textbf{(a)} Stationary solution for the one-particle distribution function $f(\theta)$ according to Eq.~\eqref{eq:UpdateRule} for a spatially homogeneous system with $L=1$, $\sigma=\sigma_0=0.5$, and $K=32$, $64$, and $128$ angular slices for various overall densities $\rho_0$ (indicated in the figure). \textbf{(b)} Representative snapshot of a stationary traveling wave pattern for $\rho_0=0.25$ and $\sigma=\sigma_0=0.5$. Color code indicates local density; length and direction of arrows indicate magnitude and orientation of local particle flow. \textbf{(c)} Phase diagram for different values of the overall density $\rho_0$ and noise amplitude $\sigma=\sigma_0$.
The analytic result for the phase boundary [refer to Eq.~\eqref{eq:rho_t}] is indicated by the dashed line.
A density segregated phase (DSP) is observed all along the phase boundary between the isotropic, spatially homogeneous phase (IHP) and the spatially homogeneous, polarized phase (SHP).
System size $L=100$, $\epsilon_x=5$, and $K=32$ for \textbf{(b)}, \textbf{(c)}.
}
\label{fig:phase_diagram_cuts}
\end{figure*}

The simple model system referred to above, Eq.~\eqref{eq:half_angle_col_rule}, has been shown to exhibit a bifurcation, which separates a parameter regime at low density / high noise, from one at high density / low noise levels. 
In the high noise (low density) regime, an isotropic, spatially homogeneous phase (IHP) is observed. This homogeneous and isotropic state can be seen as a dynamical attractor for the system's dynamics at low densities, which we will refer to as ``IHP fixed point'' in the following.
For low enough noise levels (and high enough densities), this IHP fixed point loses its stability and the system undergoes a transition toward polar order. The location of the corresponding bifurcation can be calculated directly from the Boltzmann equation, Eq.~\eqref{eq:Boltzmann_rescaled}, by performing an expansion in terms of the moments $f_k=\int_{-\pi}^{\pi}d\theta\,e^{i k \theta}f(\theta)$ of the one-particle distribution function $f(\theta)$, and studying the resulting dynamical equation for the polar order parameter $f_1$ (cf. Ref.~\cite{Bertin_short, Bertin_long} for details):
\begin{equation}
\label{eq:f1}
\partial_t f_1 = \mu_1 f_1 - \mu_2 |f_1|^2f_1 + \dots
\end{equation}
Here the various coefficients are functions of $\rho_0$, $\sigma$, and $\sigma_0$, and the dots indicate higher order and gradient terms. Although a complete description of the dynamics of $f_1$, Eq.~\eqref{eq:f1}, contains couplings to an infinite number of higher order moments $f_n$, the linear coefficient $\mu_1(\rho_0,\sigma,\sigma_0)$ can be calculated exactly within the Boltzmann equation approach and does not depend on any approximation scheme used to truncate these couplings \cite{Bertin_long,Thuroff_2013}. The condition
\begin{equation}
\mu_1(\rho_0,\sigma,\sigma_0)\bigr|_{\rho_0=\rho_t} = 0
\end{equation}
therefore provides us with an exact benchmark for the location of the bifurcation surface in parameter space. It is obtained by solving the above equation for $\rho_t$ \cite{Bertin_long}:
\begin{equation}
\label{eq:rho_t}
\rho_t = \frac{\pi}{8}\frac{1-e^{-\sigma_0^2/2}}{e^{-\sigma^2/2}-2/3}.
\end{equation}

For simplicity, we choose $\sigma=\sigma_0$ throughout this work. To test our implementation of SNAKE, we prepared different systems with varying values for the density ($0.0125\leq\rho_0\leq0.85$) and noise levels ($0.1\leq\sigma=\sigma_0\leq0.95$), starting from random initial conditions \footnote{Throughout this work, random initial conditions are chosen such that $f_n^{\alpha}=0.95+0.1\,\zeta$, where the random variable $\zeta$ is uniformly distributed on the interval $[0,1]$.}. We performed the computations on a regular rectangular spatial grid with $100\times100$ grid sites and $K=32$ angular slices, and evaluated the systems' (stationary) states after a total computation time of $T=15\,000$. The resulting phase diagram is shown in Fig.~\ref{fig:phase_diagram_cuts}(c) along with the analytical result for the bifurcation line, Eq.~\eqref{eq:rho_t}. In accordance with previous analytical \cite{Bertin_long} and numerical results \cite{Chate_long}, we observe a phase transition between an isotropic, homogeneous phase (IHP) and a broken-symmetry density segregated phase (DSP) upon crossing the bifurcation line $\rho_t(\sigma)$, Eq.~\eqref{eq:rho_t}. A typical system snapshot inside the DSP is shown in Fig.~\ref{fig:phase_diagram_cuts}(b). Deeper inside the symmetry-broken parameter domain, systems asymptotically reach a spatially homogeneous, polarized phase (SHP) after a transient episode of polar clustering and subsequent coarsening; cf. section \ref{sec:Coasening}. In the subsequent sections, we will give a more detailed analysis of the various observed phases and the corresponding transitions in between.

\subsection{\label{sec:LD_transition}First-order transition from disorder to polar order}

The transition between the isotropic homogeneous phase toward the formation of polar order (IHP $\rightarrow$ DSP) is probably one of the most studied phase transitions in active polar systems. Despite prior beliefs that this transition is of second order \cite{Vicsek}, whereby polar order was assumed to build up continuously, a growing number of more recent agent based \cite{Chate_2004,2007PhRvL..99v9601C,Chate_long} and kinetic studies \cite{Ihle_2011,Ihle:2013jg} clearly indicate that the formation of density segregated patterns at large enough system sizes render this phase transition first order.

To check for consistency of our numerical Boltzmann approach with these previous results, we studied the transition of the IHP toward polar for systems of varying linear sizes $L\in\{25,50,100\}$, subject to two different noise levels $\sigma=\sigma_0\in\{0.35,0.6\}$. For each choice of the system size and noise level, we prepared a set of systems at different overall densities $\rho_0\in[0.05<\rho_t(\sigma),\,0.64>\rho_t(\sigma)]$ using random initial conditions. To quantify the phase transition, we computed the local polar order parameter $\bm\varphi_{\alpha}$ from the numerical solution of the \emph{density matrix} $f_n^{\alpha}$,
\begin{equation}
\bm\varphi_{\alpha} \equiv \frac{2\pi}{K}\,\sum_{n=1}^Ke^{i \theta_n}f_n^{\alpha}.
\end{equation}
From this, we then computed the spatially averaged polar order parameter,
\begin{equation}
\langle\bm\varphi\rangle_x\equiv\frac{1}{L^2}\sum_{\alpha=1}^{L^2}\bm\varphi_{\alpha},
\end{equation}
after the systems have settled into a stationary state (which, here and in the following, shall include moving patterns of ``stationary'' shape, as observed inside the DSP).

\setlength{\tabcolsep}{-3.4em}
\begin{figure}
\centering
\begin{tabular}{cc}
\includegraphics{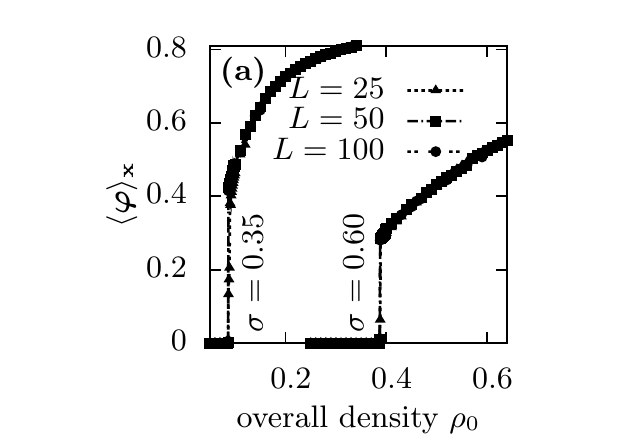} & \includegraphics{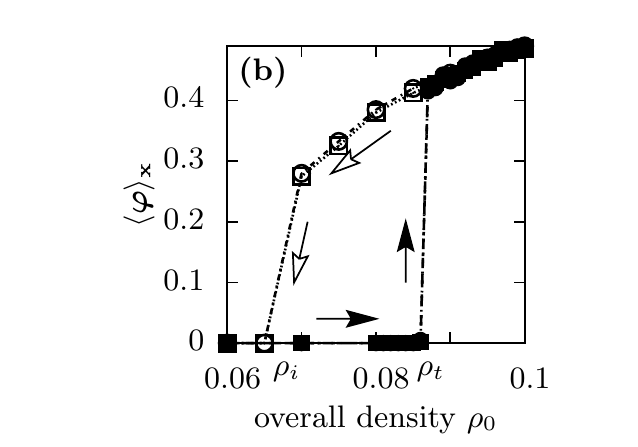}
\end{tabular}
\caption{
\textbf{(a)} Average momentum against density $\rho$ for two noise levels $\sigma=\sigma_0$ and three system sizes $L$; specific values are indicated in the figure [$\epsilon_x=5$]. The results indicate that the transition is first-order. \textbf{(b)} Hysteresis study. Average momentum against overall density $\rho_0$ for $\sigma=\sigma_0=0.35$ and two system sizes $L\in\{50,100\}$. Solid symbols (path indicated by solid arrows) indicate the system's stationary state when starting from random initial conditions. Open symbols (path indicated by open arrows) indicate the system's stationary states upon lowering the density ``quasi-statically'' starting from a stationary wave-like pattern at $\rho_0>\rho_t$. The system exhibits hysteresis, corroborating the assertion that the underlying ordering transition is first order.
}
\label{fig:LD_hysteresis}
\end{figure}

The results of this study are shown in Fig.~\ref{fig:LD_hysteresis}(a). Our results indicate a jump discontinuity of the polar order parameter $\langle\bm\varphi\rangle_x$ at the transition to polar order, which is accompanied by the formation of spatial heterogeneities, typically in the form of traveling wave patterns (cf. Fig.~\ref{fig:phase_diagram_cuts}(b); see also section \ref{sec:Cluster}). To within our chosen density resolution, $\Delta\rho=0.01$, the size of the observed jump-discontinuity is virtually independent of the system size for $L\gtrsim50$, but depends on the noise levels in the system. For the smallest system sizes considered, $L=25$ and smaller (not shown), the size of the jump-discontinuity shrinks down to zero, eventually rendering the observed phase transition continuous for system sizes below a critical size $L^*$. These observations are in agreement with previous studies on this phase transition in the context of the Vicsek model \cite{Chate_2004,Chate_long} and Enskog-like kinetic theories \cite{Ihle_2011,Ihle:2013jg}. For a more detailed discussion of the critical system size $L^*$, we refer the reader to Refs.~\cite{Ihle_2011,Ihle:2013jg}.

To scrutinize the subcritical nature of the bifurcation leading to the formation of polar order in large enough systems, we further checked for the existence of hysteresis effects in this model system. To this end, we numerically computed the ``stationary'' solution of the \emph{density matrix} $f_n^{\alpha}$ inside the DSP, and then quasi-statically reduce the overall density $\rho_0$ in small steps. Here and in the following, the term ``quasi-static'' change of the overall density refers to the fact that the system is given a sufficient amount of time to equilibrate in between successive adjustments in $\rho_0$. Two typical outcomes of such hysteresis experiments are shown in Fig.~\ref{fig:LD_hysteresis}(b) for two different system sizes. There, closed symbols and arrows indicate the path a system follows in the $\rho_0$--$\langle\bm\varphi\rangle_x$ plane, when starting from random initial conditions, exhibiting the familiar jump discontinuity at $\rho_0=\rho_t$. Open symbols and arrows indicate the inverse situation, where systems evolve from an initial state inside the ordered DSP parameter regime. In this latter case, our results indicate that the corresponding dynamical fixed point is stable down to values of $\rho_0=\rho_{\text{i}}<\rho_t$ well below the actual transition density $\rho_t$, thus corroborating the subcritical nature of the corresponding dynamical bifurcation. Similar hysteresis effects have been reported in Refs.~\cite{Chate_long,Peshkov:2012tu,Ihle:2013jg}.

\subsection{\label{sec:DSP}Density segregated polar phase}

\setlength{\tabcolsep}{-1.5em}
\begin{figure*}
\begin{tabular}{ccc}
\includegraphics{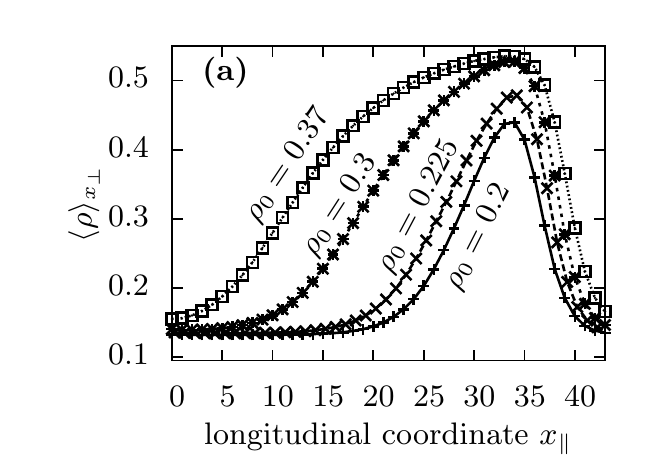} & \includegraphics{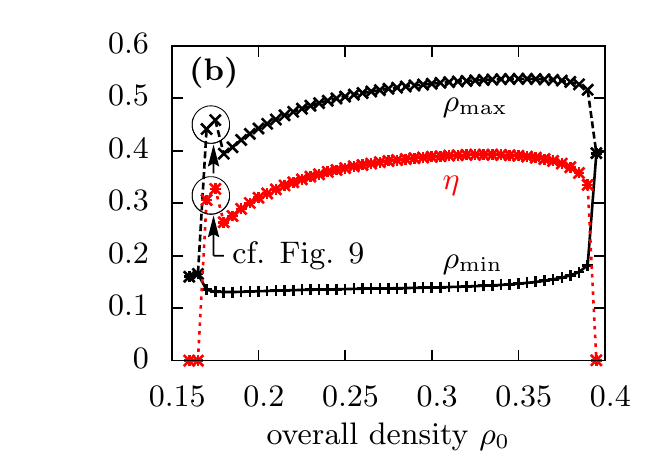} & \includegraphics{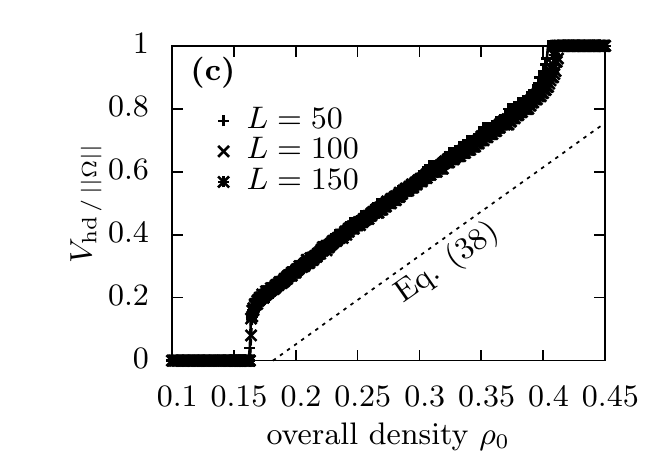}
\end{tabular}
\caption{
Density variation inside the DSP parameter regime ($\sigma=\sigma_0=0.5$).
\textbf{(a)} Typical profiles of wave-like patterns inside the density segregated phase for various overall densities $\rho_0$. The asymmetry of the wave profiles increases with density ($\rho_t\approx0.21$ for the noise values considered). \textbf{(b)} Density segregation as a function of the overall density $\rho_0$. Data points have been obtained by quasi-statically changing $\rho_0$ starting at a stationary traveling wave pattern at $\rho_0=0.25$. As a consequence of hysteresis effects [cf. Figs. \ref{fig:LD_hysteresis}(b), \ref{fig:HD_hysteresis}], phase space coexistence can be observed between the DSP and the IHP at low densities, as well as between the DSP and the SHP at higher densities. \textbf{(c)} Surface fraction of the HD phase as function of the overall density $\rho_0$. Within the DSP, the surface fraction varies virtually linear with $\rho_0$ in agreement with the mean field picture, Eq.~\eqref{eq:MFpic}. At the transitions IHP $\rightarrow$ DSP and DSP $\rightarrow$ SHP, the surface fraction exhibits a discontinuous jump.
}
\label{fig:density_segregation}
\end{figure*}

Since we have chosen to numerically solve the Boltzmann equation in real space, rather than in Fourier space, we expect \SNAKE~to be applicable even for parameters well inside the ordered phase, where Fourier space approaches would have to include an exceedingly large number of modes.
Therefore, we continue our study of polar systems by presenting a more detailed analysis of the DSP region in parameter space. In anticipation of our discussion in section \ref{sec:Cluster}, we note that we frequently observe two distinct types of patterns within the DSP: traveling wave patterns and ``\emph{cluster lane patterns}''. While the former pattern shape is by now well known to occur in polar active systems with metric interactions \cite{Chate_long,Bertin_long,Ihle:2013jg,Thuroff_2013}, to our knowledge the emergence of \emph{cluster lane patterns} has not yet been reported in the literature so far. We devoted the separate section \ref{sec:Cluster} to these novel patterns, and restrict our present discussions to the well-known case of traveling waves.

After initial transients, during which ``polar droplets'' form and coalesce out of random initial conditions, the resulting ``stationary'' traveling wave patterns (when viewed in a co-moving frame) become translationally invariant in the lateral direction $x_{\perp}$ (perpendicular to the wave's velocity vector). The structure of these wave patterns can, therefore, be described by  means of the longitudinal coordinate $x_{\|}$ (along the direction of the wave's velocity vector). Figure \ref{fig:density_segregation}(a) shows typical density profiles of such polar waves as a function of the distance from the transition line at $\rho_t(\sigma)$. The observed wave profiles exhibit a characteristic front-rear asymmetry, similar to those reported in previous numerical studies \cite{Chate_long}, whereby the asymmetry becomes more pronounced as the parameters are chosen deeper inside the ordered region.

The waves themselves can be viewed as high-density structures, traveling on an isotropic, low-density background [cf. appendix~\ref{app:waves} and Fig.~\ref{fig:momentum_profiles}]. To quantify the density segregation between the wave fronts and the isotropic background as function of the distance from the system's phase boundary, we introduce the following characteristic density scales:
\begin{subequations}
\begin{eqnarray}
\rho_{\text{max}} &\equiv& \max_{x_{\|}}\,\langle\rho(\vec x)\rangle_{x_{\perp}}, \\
\rho_{\text{min}} &\equiv& \min_{x_{\|}}\,\langle\rho(\vec x)\rangle_{x_{\perp}},
\end{eqnarray}
as measures for the density levels within the high-density, moving phase (``HD phase''), and the low-density, isotropic background (``LD phase''), respectively.
Figure \ref{fig:density_segregation}(b) shows the characteristic density scales, and the density separation parameter,
\begin{equation}
\eta\equiv\rho_{\text{max}}-\rho_{\text{min}},
\end{equation}
\end{subequations}
as a function of the system's overall density $\rho_0$ at fixed noise levels. The data shown in Fig.~\ref{fig:density_segregation} has been computed using the following simulation protocol: For a system well inside the DSP parameter regime, we followed its time evolution starting from a random initial condition until it reached a stationary state with a traveling wave pattern. The density \emph{density matrix} $f_n^{\alpha}$ corresponding to this stationary wave pattern was then used as initial condition in subsequent numerical computations, during which the density level was ``quasi-statically'' decreased or increased until the density segregation parameter dropped back to zero, i.e. until a spatially homogeneous state is reached.

This protocol allows us to explore the parameter range over which the ``DSP fixed point'' of the system's dynamics remains \emph{locally} attracting. Before we embark on a more detailed discussion on the stability of the various dynamical fixed points and observed phase coexistence in parameter space in section \ref{sec:HD_transition}, we take a little detour and briefly comment on the observed wave profiles as the overall density is varied. First, we observe that the density level of the LD phase, $\rho_{\text{min}}$, is virtually independent of the overall density $\rho_0$ across the entire region of existence of the DSP. Any ``surplus density'' arising from an increase in $\rho_0$ must therefore be accommodated within the HD phase, and, consequently, implies a change in the wave profiles.
At the lowest densities ($\rho_0<\rho_t$) where wave patterns are still stable, we observe wave profiles of relatively weakly pronounced front-rear asymmetry [cf. Fig.~\ref{fig:density_segregation}(a), $\rho_0=0.2$]. As the overall density is increased slightly above $\rho_t$, surplus density is being accommodated primarily by increasing the characteristic density $\rho_{\text{max}}$, leaving the front-rear asymmetry virtually unchanged [cf. Fig.~\ref{fig:density_segregation}(1), $\rho_0=0.225$]. For still larger values of the overall $\rho_0$, the characteristic density $\rho_{\text{max}}$ quickly saturates and surplus density is being accommodated by increasing the width of the solitary waves. This, in turn, leads to a drastic increase in front-rear asymmetry.

The latter regime of wave broadening can be understood intuitively by employing a simple mean field picture which assumes constant density levels within the HD and LD phases, respectively. To this end, consider a system of arbitrary volume $||\Omega||$ in the DSP wave broadening parameter regime. Since both characteristic densities, $\rho_{\text{min}}$ and $\rho_{\text{max}}$, are constant by assumption, conservation of particle number $N\simeq V_{\text{hd}}\,\rho_{\text{max}}+V_{\text{ld}}\,\rho_{\text{min}}$ ($V_{\text{hd/ld}}$: volumes occupied by HD and LD phases, respectively) dictates:
\begin{equation}
\label{eq:MFpic}
\frac{V_{\text{hd}}}{||\Omega||}=\frac{\rho_0-\rho_{\text{min}}}{\eta}.
\end{equation}
Figure \ref{fig:density_segregation}(c), where the surface fraction $V_{\text{hd}}/||\Omega||$ is recorded as a function of $\rho_0$ confirms this linear dependence, and thus corroborates the above mean field picture. Moreover, since $V_{\text{hd}}\leq||\Omega||$, Eq.~\eqref{eq:MFpic} implies a crossover scale $\rho_{h}=\mathcal{O}(\rho_{\text{max}})$, above which the LD phase gets depleted and the system becomes spatially homogeneous.
We checked that the mean field picture conveyed by Eq.~\eqref{eq:MFpic} is actually independent of the system size $L$, and becomes quantitatively robust against changes in the system size for $L\gtrsim50$; cf. Fig.~\ref{fig:density_segregation}(c).
Interestingly, the composition of the system in terms of HD and LD phases as a function of of the overall density $\rho_0$ exhibits qualitatively analogous behavior in the two-dimensional lattice model of active Ising spins of Ref. \cite{2013PhRvL.111g8101S}, where a $Z_2$ (up-down) symmetry is spontaneously broken. This observation strongly suggests that the qualitative features of the flocking transition toward spatially homogeneous polar order in active systems might actually be universal across different symmetry classes.

\subsection{\label{sec:HD_transition}Transition to spatially homogeneous polar state and pattern selection}

\setlength{\tabcolsep}{-2em}
\begin{figure}
\centering
\includegraphics{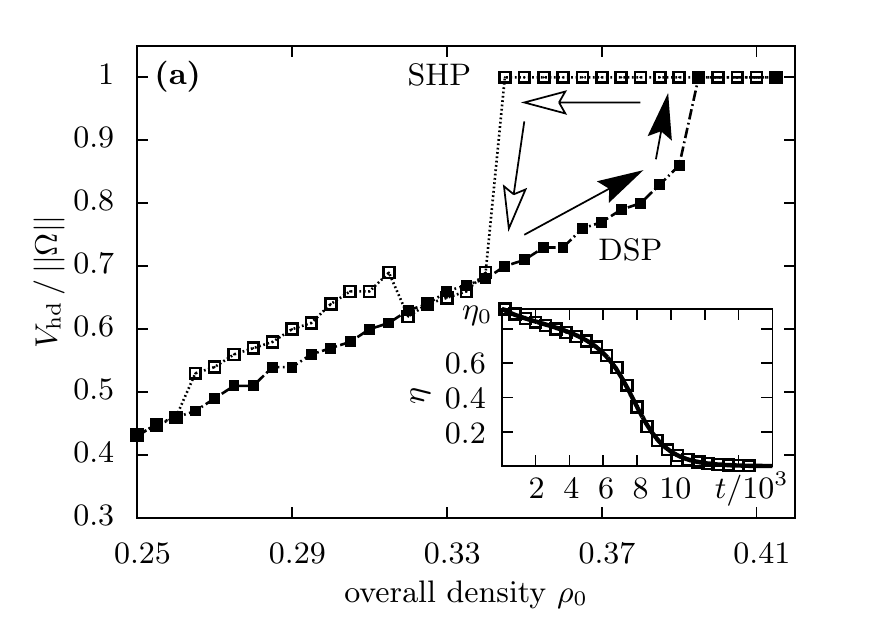}\\
\vspace{.2cm}
\includegraphics{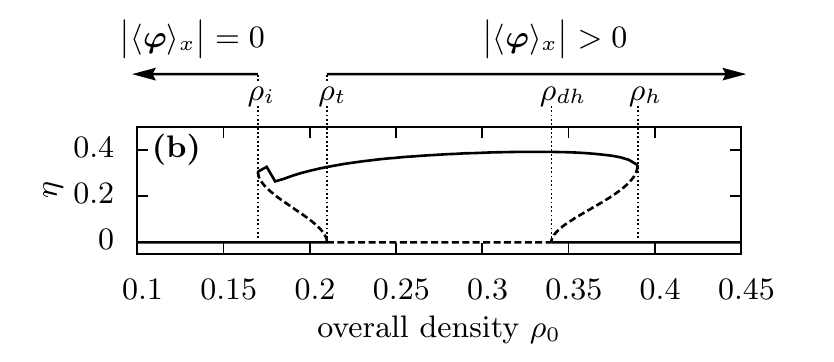}\\
\caption{
\textbf{(a)} Hysteresis study ($\sigma=\sigma_0=0.25$). Surface fraction of the HD phase against overall density $\rho_0$. Solid symbols (path indicated by solid arrows) indicate the system's stationary state upon increasing $\rho_0$ quasi-statically, starting at a wave-like pattern at $\rho_0=0.25$. Open symbols (path indicated by open arrows) indicate the system's stationary states upon quenching the density to successively lower values, starting from a stationary, spatially homogeneous state computed at $\rho_0=0.415$. The system exhibits hysteresis, corroborating the assertion that the underlying ordering transition is first order. \textit{Inset:} Dynamics of the density segregation parameter $\eta$ (open data points) follows the subcritical bifurcation model, Eq.~\eqref{eq:subcrit_bifircation} (solid line). 
\textbf{(b)} Sketch of the bifurcation diagram for the bifurcation parameter $\rho_0$ (at $\sigma=\sigma_0=0.5$). Solid lines indicate stable fixed points of the system's dynamics and have been determined numerically. Dashed lines are drawn ``by hand'' and indicate the estimated position of unstable fixed points.
}
\label{fig:HD_hysteresis}
\end{figure}

In the previous section, we investigated the evolution of systems inside the DSP regime, as the overall density $\rho_0$ is varied. We observed that the DSP parameter regime, which can be defined by non-zero values of the density separation parameter $\eta$,  is hallmarked by a linear variation of the surface fraction of the HD phase, $V_{\text{hd}}/||\Omega||$, with $\rho_0$. Importantly, the DSP regime is delimited by sharp jumps in $\eta$ and $V_{\text{hd}}/||\Omega||$ at both of its boundaries, i.e. at the transition toward spatially homogeneous isotropic state (low densities) and at the transition toward a spatially homogeneous polar ordered state (high densities). As has been discussed before (e.g. Refs.~\cite{Chate_long,Ihle:2013jg} and section \ref{sec:LD_transition}), such non-regular behavior at the low density limit of the DSP regime can be attributed to the first-order nature of the ordering transition between the IHP and DSP parameter regimes. However, for the transition between the DSP regime and the SHP regime comparable results are not available, yet. The present section is devoted to fill this gap, and to illuminate the nature of the phase transition between DSP and SHP regimes. Similar to the case of the IHP $\leftrightarrow$ DSP phase transition, we will observe hysteresis effects which are highly suggestive of a first-order transition between the DSP and SHP regimes.

To assess the nature of the DSP $\leftrightarrow$ SHP transition, we chose density and noise parameters inside the SHP regime ($\rho_0=0.415$, $\sigma=\sigma_0=0.5$), and let the system evolve until it has reached a state of spatially homogeneous polar order. The corresponding stationary solution for the \emph{density matrix} $f_n^{\alpha}$ was then used as initial condition in subsequent computations.
 In contrast to the computation protocol in section \ref{sec:LD_transition} (where parameters were adjusted quasi-statically), the system's overall density $\rho_0$ was then quenched to successively lower values $\rho_0\in\{0.41,\,0.405,\dots,\,0.25\}$, and the resulting pattern at asymptotically long times was recorded; for details of the numerical protocol cf. Appendix \ref{sec:numerical_protocol_hd_hyst}. 

In Fig.~\ref{fig:HD_hysteresis}(a) the corresponding results of this study are shown in terms of the surface fraction $V_{\text{hd}} / ||\Omega||$ occupied by the HD phase. There, open symbols and arrows indicate data points corresponding to initial conditions inside the SHP regime. Closed data points refer to initial conditions inside the DSP regime (where the overall density was then quasi-statically increased). 
The data show that the transition between the DSP and SHP regimes is accompanied by hysteresis effects, supporting our initial claim that the corresponding transition is first-order.
Phase space coexistence between both regimes can be observed for densities $\rho_{dh}(\sigma=\sigma_0=0.5)\approx0.345\leq\rho_0\leq\rho_h(\sigma=\sigma_0=0.5)\approx0.39$ \footnote{The small gap between the data points in Fig.~\ref{fig:HD_hysteresis}(a) for $0.265\leq\rho_0\leq0.295$ is due to a different pattern selection of the system on both branches of the hysteresis loop: While the solid branch refers to a single wave band wrapped around the donut, the density quench represented by the open symbols in the corresponding density range results in formation of a two-banded patters and thus occupies a slightly larger fraction of the overall volume.}.
Further evidence confirming the first-order nature of this transition comes from an analysis of the dynamics of the density separation parameter $\eta$ at the point where the actual transition DSP $\rightarrow$ SHP occurs [solid branch of the hysteresis loop in Fig.~\ref{fig:HD_hysteresis}(a); $\rho_0=0.395$]. In the inset of Fig.~\ref{fig:HD_hysteresis}(a), we show the dynamical evolution $\eta(t)$ (open data points), together with a numerically computed solution to a model equation for a subcritical bifurcation,
\begin{equation}
\label{eq:subcrit_bifircation}
\begin{split}
\dot{\eta} = 0.15675\,\eta &+ 0.424\,\eta^3 - 0.3\,\eta^5,\\
\eta(0) &= \eta_0,
\end{split}
\end{equation}
and find excellent agreement.

Summarizing the results from section \ref{sec:Applications}, we can now extend the picture conveyed by the phase space diagram in Fig.~\ref{fig:phase_diagram_cuts}(c), which has been recorded using random initial conditions at different points in parameter space. Our discussions in this and previous sections disclosed regions in parameter space, where phase selection by the system is not uniquely related to the specific choice of parameters (i.e. density and noise levels). Instead, phase selection is sensitive to initial conditions in these parameter regions, and IHP / DSP coexistence at the low-density end, and DSP / SHP coexistence at the high-density end of the DSP regime is observed. We condensed these findings by sketching the corresponding bifurcation diagram in the $\rho_0$--$\eta$ plane in Fig.~\ref{fig:HD_hysteresis}(b). There, solid lines indicate the position of stable fixed points of the system's nonlinear dynamics, as measured by actual computations. Dashed lines are drawn ``by hand'' and indicate an estimate of the position of the corresponding unstable fixed points. Figure \ref{fig:HD_hysteresis}(b) corresponds to a horizontal cut through the phase diagram shown in Fig.~\ref{fig:phase_diagram_cuts}(c) at $\sigma=\sigma_0=0.5$. Comparing both figures, we see that the phase diagram in Fig.~\ref{fig:phase_diagram_cuts}(c) displays the DSP only in the relatively narrow window $\rho_t\lesssim\rho_0\lesssim\rho_{dh}$ since initial conditions very close to the $\eta=0$ fixed point are chosen. Due to the subcritical character of both transitions, IHP $\leftrightarrow$ DSP and DSP $\leftrightarrow$ SHP, density segregated patterns can actually be observed over the much broader parameter range $\rho_i<\rho_0<\rho_{h}$. Interestingly, an analogous extension of the existence region of density segregated patterns has recently been reported in the context of nematic ordered active systems \cite{Peshkov:2012tu,1367-2630-15-8-085032}.

\subsection{\label{sec:Cluster}Phase separated patterns}

\setlength{\tabcolsep}{0em}
\begin{figure*}
\centering
\begin{tabular}{ccc}
\includegraphics{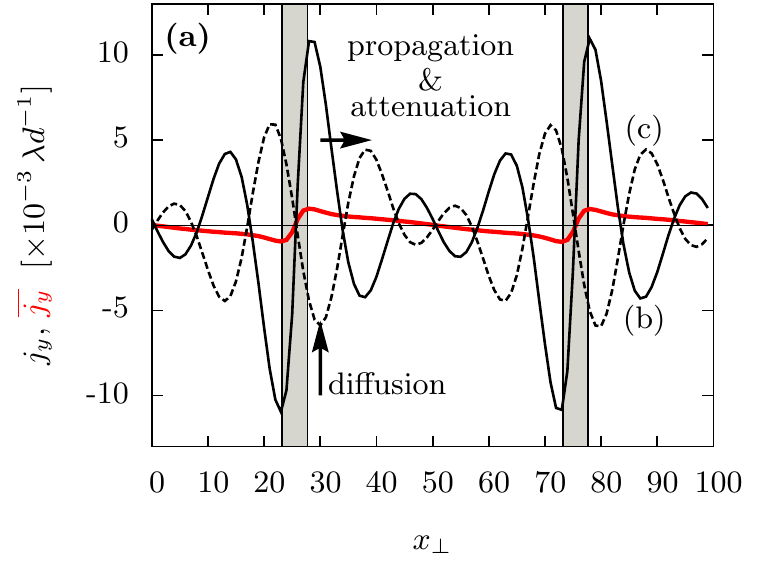} & \includegraphics{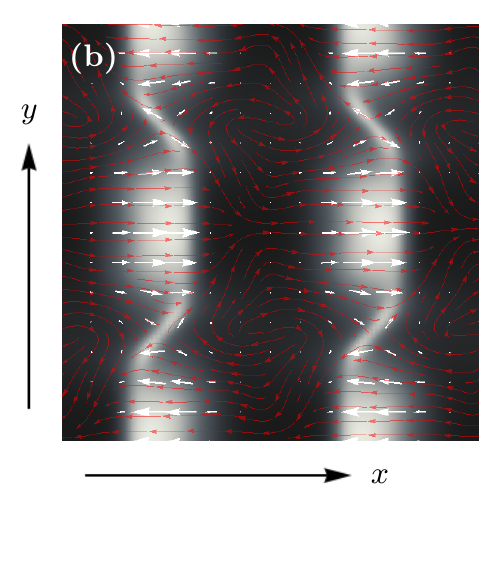} & \includegraphics{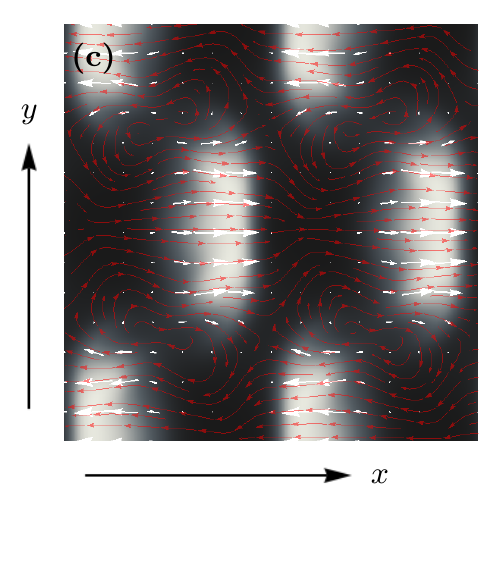}
\end{tabular}
\vspace{-.3cm}
\caption{Cluster lane patterns. \textbf{(a)} Particle currents in the cluster lane solution of the active Boltzmann equation. Black curves: Instantaneous realizations of the longitudinally averaged, transverse particle currents $j_{y}(y,t)$ as functions of the (transverse) $y$ coordinate taken at different instances of time; cf. snapshots in (b,c). Red curve: Time averaged transverse particle currents $\overline{j_{y}}(y)$. The gray shaded areas indicate the \emph{depletion zones} where cluster grazing occurs ``feeding'' the \emph{cluster zones} (white areas). \textbf{(b)} Snapshot of a cluster lane patten taken at the instant $t_1$ where the clusters of both lanes collide. \textbf{(c)} Snapshot of the same cluster lane patten taken at some later instant $t_2>t_1$ where the clusters of both lanes spread due to transverse diffusion. \textbf{(b,c)} Color code indicates local density levels, white arrows indicate magnitude and direction of local momentum, red stream lines indicate macroscopic flow patterns.
}
\label{fig:cluster_particle_currents}
\end{figure*}

In previous works on polar systems with identical \cite{Bertin_long}, or similar model properties \cite{Chate_long,Ihle:2013jg}, the emergence of density segregated states has been discussed in terms of traveling wave patterns with ``infinite'' extent in the lateral direction \cite{2014arXiv1401.1315C}; cf. snapshot in Fig.~\ref{fig:phase_diagram_cuts}(b). However, in our numerical studies we find an additional, genuinely distinct type of density segregated patterns, which is fundamentally different from the familiar solitary wave patterns. These patterns consist of a number of regularly shaped, moving high density regions of \emph{finite lateral extent}, where each such region carries a large net polarity. To emphasize their finite spatial extent, we will refer to these high-density polar regions as \emph{(polar) ``clusters''}. Intriguingly, the polarities of these clusters are oriented along a common broken-symmetry axis with orientations distributed such that the overall symmetry of the system is nematic. More precisely, clusters are arranged in \emph{lanes}, where each lane hosts a train of equally spaced clusters of parallel polarities. In our numerical studies, we observe two distinct types of \emph{cluster lane patterns}: First, we find antisymmetric cluster lane solutions, where all cluster lanes have identical geometrical properties (same lane width and number and shape of polar clusters), except for a reflection of the cluster polarity from one lane to the next, thus rendering the overall symmetry of the system purely nematic. Typical snapshots of this first type of antisymmetric cluster lane patterns are shown in Figs.~\ref{fig:cluster_particle_currents}(b,c) and in Fig.~\ref{fig:cluster_lane_geometry}(b).  Second, we observe asymmetric cluster lane patterns where neither the number of clusters per lane, nor the morphological properties of clusters in different lanes are symmetric; cf. Fig.~\ref{fig:cluster_lane_geometry}(c). We note that the lane widths of this latter type of asymmetrical clusters were subject to a very slow but continuous drift during the computation times considered in this work. We can, therefore, not exclude the possibility that this asymmetrical pattern would actually give rise to the more familiar wave pattern in the limit of long times (in which case the asymmetrical pattern would constitute an intermediate asymptote of the system's dynamics). In contrast, we find that the structural properties of the antisymmetric cluster lane patterns do converge, which strongly suggest that these patterns constitute a genuine limit-cycle solution of the underlying Boltzmann equation, Eq.~\eqref{eq:Boltzmann_rescaled}. In the following we will, therefore, restrict ourselves to examine antisymmetric cluster lane patterns which, moreover,  are more frequently observed than their asymmetric counterparts. In particular, we give a detailed discussion highlighting the basic mechanisms stabilizing these remarkable patterns.

Before we embark on a more thorough discussion on the basic mechanisms stabilizing these nematic ordered patterns, we briefly comment on the observation statistics of antisymmetric cluster lane patterns in our numerical solutions. As discussed in detail in Appendix~\ref{sec:cluster_lane_orientations}, the broken symmetry axis in such cluster lane patterns is effectively constrained to one of the two coordinate directions, due to the finite size of the computational grids used in our numerical studies. To obtain a counting statistics, we chose different sets of model parameters inside the DSP regime and used 350 different seeds per set of parameters to prepare independent systems with random initial conditions. We then let each of these systems evolve until a structurally stable, density segregated state had developed. From all 350 patterns, we selected those with a mean polar / nematic order along one of the coordinate axes and computed the observation frequency $p$ of antisymmetric cluster lane patterns among these ``parallel patterns'' \footnote{For the largest system sizes considered ($L=150$), approximately 50-60 seeds out of a total of 350 seeds lead to parallel patterns.}. While the observation frequency of antisymmetric cluster lane patterns is rather low for the smallest systems sizes considered in this work ($L=50$; $p\approx5\%$), these patterns make up a considerable fraction of all parallel patterns for system sizes $L\gtrsim100$, in which case we find $p\approx25\%$ (we find $p\approx40\%$ if we count all cluster lane patterns, including asymmetric cluster lanes).

We shall now give a more precise definition of the term ``cluster lane'' and elucidate the dynamical mechanisms underlying their formation and stabilization. To this end, consider the antisymmetric cluster lane pattern of Figs.~\ref{fig:cluster_particle_currents}(b,c) with axis of nematic order along the $x$-direction. To assess the particle exchange dynamics between the various clusters, we recorded the transverse particle currents $g_y(\vec x,t)$, Eq.~\eqref{eq:def_momentum}, and computed its longitudinal and time average according to
\begin{subequations}
\begin{eqnarray}
j_y(y,t) &\equiv& \frac{1}{L}\,\int_0^Ldx\,g_y(\vec x,t),\\
\overline{j_y}(y) &\equiv& \frac{1}{\Delta T}\,\int_{t_0}^{t_0+\Delta T}dt\,j_y(y,t),
\end{eqnarray}
\end{subequations}
where the time scales $t_0=\Delta T=10\,000\,\lambda^{-1}$ are chosen such that the computation of the time averaged particle currents is done on the basis of a structurally stable cluster lane pattern, i.e. in the limit of long times. Fig.~\ref{fig:cluster_particle_currents}(a) shows $\overline{j_y}(y)$ along with the (instantaneous) particle currents $j_y(y,t)$ corresponding to two particular time instances, indicated by the snapshots in Fig.~\ref{fig:cluster_particle_currents}(b,c).

From inspection of the temporal average of the particle current profile $\overline{j_y}(y)$ [red curve in Fig.~\ref{fig:cluster_particle_currents}(a)], we extract the following structural features: The overall functional form of $\overline{j_y}(y)$ resembles a sawtooth shape with large positive slopes (i.e. divergence of $\overline{j_y}$) in narrow \emph{depletion zones} close to the cluster lane boundaries, and a shallow negative slope (i.e. convergence of $\overline{j_y}$) throughout the spatially extended \emph{cluster zones}. The borders $y=y_b^{i}$ (with $i=1,2,\dots$ labeling the different border lines) delineating adjacent cluster lanes are located at the zeros of the particle current inside the \emph{depletion zones}, for which
\begin{equation}
j_y^{(s)}(y,t)\bigr|_{y=y_b^{i}}=0\quad\text{for all } t \text{ and } i
\end{equation}
holds true in the stationary state (indicated by the superscript $s$). The sawtooth shape of the time averaged current $\overline{j_y}$ therefore confers the following picture illustrating the balance of particle exchanges across different cluster lanes: The bulk of the clusters, i.e. the \emph{cluster zones} inside each lane, receive particles at the expense of a narrow \emph{depletion zone} at the cluster lane borders. Within these \emph{depletion zones}, large transverse, inwardly directed particle currents, ``feeding'' the bulk of the clusters, arise due to periodically recurring ``cluster grazing''; cf. solid black curve in Fig.~\ref{fig:cluster_particle_currents}(a) and snapshot in Fig.~\ref{fig:cluster_particle_currents}(b). As time progresses, these particle currents propagate toward the center of the clusters. Concomitantly, rotational diffusion attenuates these propagative modes and gives rise to net diffusive currents at the tails of these modes which entail diffusive spreading of each cluster across the lane boundaries; cf. dashed black curve in Fig.~\ref{fig:cluster_particle_currents}(a) and snapshot in Fig.~\ref{fig:cluster_particle_currents}(c). In steady state, these diffusive currents balance such that there is no net current across the lane boundaries. When the inwardly directed (attenuated) transverse currents eventually arrive at the cluster centers (which, again, correspond to zeros of the transverse particle currents for all $t$), the aligning collision rule converts them into longitudinal currents, thus establishing a feedback mechanism to the macroscopic order parameter.

\begin{figure}[t]
\centering
\includegraphics{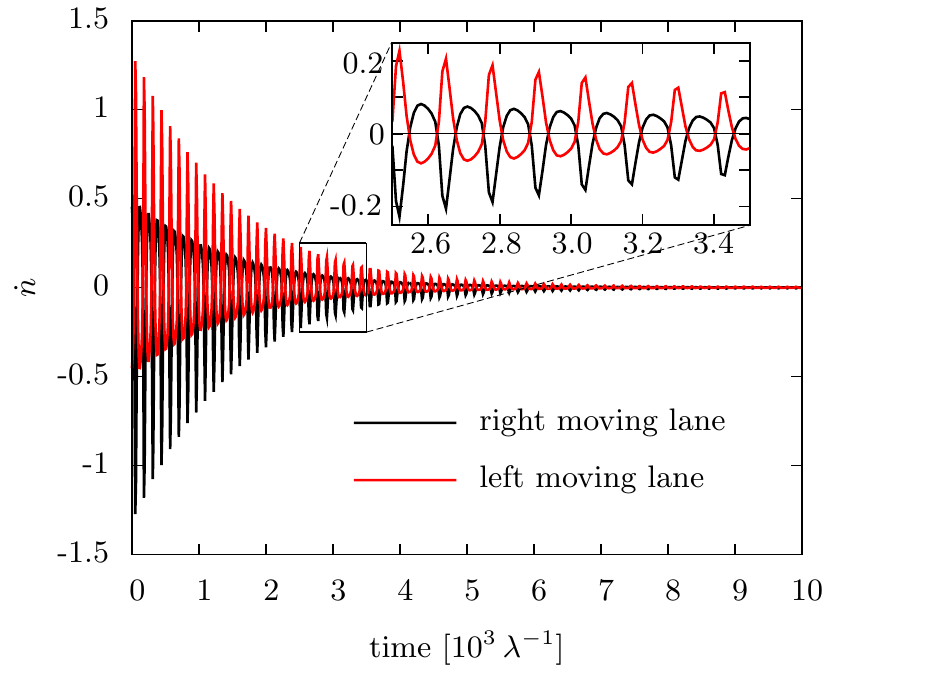}
\caption{Particle exchange across the cluster lane boundaries in the long time limit. In the plots, $t=0$ corresponds to a simulation time of $t_{\text{sim}}=10\,000\,\lambda^{-1}$, during which the cluster lane pattern has developed from random initial conditions at $t_{\text{sim}}=0$. Particle exchanges across the cluster lane boundaries are driven by cluster grazing (spikes) and lateral diffusion of clusters (bumps); see \emph{inset}. In the long time limit, the magnitude of the instantaneous value of the net particle exchange $\dot{n}$ across the cluster lane boundaries converges to zero.}
\label{fig:cluster_particle_exchange}
\end{figure}

To demonstrate the structural stability of these antisymmetric cluster lane patterns, we further computed the net particle exchanges [$\delta C$: boundary delineating the right moving cluster in Figs.~\ref{fig:cluster_particle_currents}(b,c)]
\begin{equation}
\dot{n}=\oint_{\delta C}d\vec x\,g_y(\vec x,t)=L\,\left[j_y(y_b^1,t)-j_y(y_b^2,t)\right]
\end{equation}
across the cluster lane boundaries at $y_b^2,\,y_b^1$ ($y_b^2>y_b^1$). Since in the stationary state $j_y^{(s)}(y_b^i,t)=0$, we expect the magnitude of $j_y(y_b^i,t)$ to converge to zero in the limit $t\rightarrow\infty$. In Fig.~\ref{fig:cluster_particle_exchange}, the net rates $\dot{n}$ at which the left and right moving cluster lanes in Figs.~\ref{fig:cluster_particle_currents}(b,c) gain and loose particles are plotted as a function of time in the long time regime $\lambda\,t\in[10\,000,20\,000]$, where the cluster lane pattern is virtually stationary. Due to overall particle conservation, the red and black curves in Fig.~\ref{fig:cluster_particle_exchange} sum up to zero. The net rate at which both cluster lanes gain and lose particles are extremely small in the displayed long time limit and decay exponentially as $t\rightarrow\infty$. This indicates that a balance of particle numbers is being established between both lanes, such that the corresponding cross-lane particle currents cancel in the limit of asymptotically long times. The closeup view in the inset of Fig.~\ref{fig:cluster_particle_exchange} reveals that this balancing process is driven by cluster grazing events (spikes) during which the right moving lane gives off particles to the left moving lane, and transverse diffusion (bumps) which leads to a net current of particles into the right moving cluster lane. We stress, however, that even in the stationary state large amounts of particles are transported laterally across the cluster lane boundaries, and only the net exchange of particles vanishes due to a global cancellation of periodically oscillating particle currents which are constantly maintained by cluster grazing and transverse diffusion.

In summary, cluster lane patterns constitute an intriguing, stable limit-cycle solution of the Boltzmann equation. The transport processes entailed by these cluster lane patterns differ from those of the more familiar solitary wave patterns in two important respects: First of all, in contrast to the solitary wave solution, where a macroscopic transport of particles is confined to the longitudinal direction, the antisymmetric cluster lane pattern approaches a stationary state in which macroscopic, periodically oscillating particle currents give rise to an interesting local net transport of particles in lateral directions, which is balanced only on global scales. Second, the global order of this limit-cycle solution is nematic rather than polar. The cluster lane solution of the Boltzmann equation thus highlights a physical mechanism by which a system of self-propelled polar particles and purely polar interactions is capable of eliciting a macroscopic state of global nematic order. Put differently, clusters could be viewed as polar quasi-particles with only nematic interactions, thus rendering the system nematic on macroscopic scales. For a sample movie of a cluster lane solutions refer to the Supplemental Material \cite{Supplement}.

Before we conclude, some additional remarks are in order. Our numerical results, presented in this section, provide strong evidence for the actual stability of these novel cluster lane patterns. Moreover, we checked that the emergence of cluster lane patterns is not a singular property of the specific features of the BDG model: As a matter of fact, we observed the formation of cluster lane patterns over a large range of parameters for a class of more general collision rules (cf. Ref.~\cite{Thuroff_2013}) deviating from the half-angle alignment rule, Eq.~\eqref{eq:half_angle_col_rule}, employed in the BDG model (data not shown). This suggests that the cluster lane solution is of an even more general scope within the framework of the Boltzmann equation approach to active polar systems. Nevertheless, it is important to keep in mind that the Boltzmann equation itself rests on a number of specific assumptions, including the absence of multi-particle interactions beyond binary collisions and the neglect of correlation effects among particle states as a consequence of the molecular chaos assumption. To which extent the emergence of cluster lane patterns is hinged on such hidden assumption constitutes an important question to be addressed in order to clarify on the role of such cluster lane patterns for actual active matter systems. We, therefore, hope that future studies employing alternative modeling approaches will shed light on this important question.

\subsection{\label{sec:Coasening}Coarsening dynamics}

 \setlength{\tabcolsep}{-3.4em}
\begin{figure}
\centering
\begin{tabular}{cc}
\includegraphics{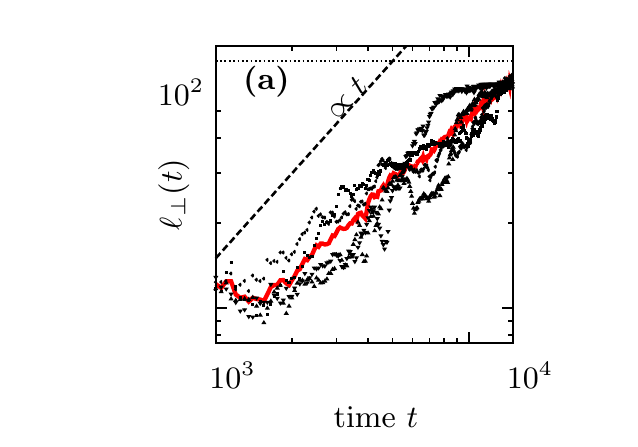} & \includegraphics{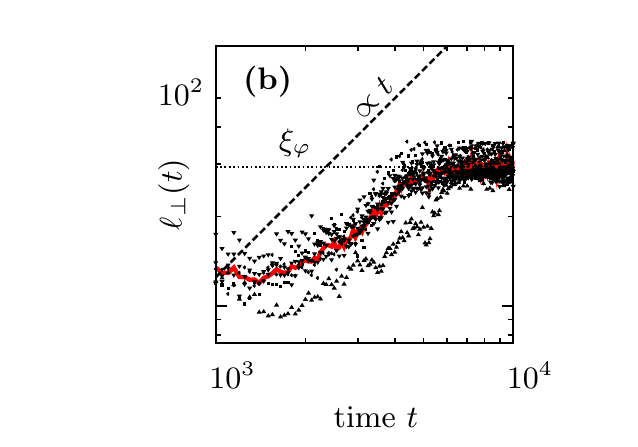}
\end{tabular}
\caption{
Scaling of the coarsening dynamics inside the DSP regime at intermediate and long times. \textbf{(a)} Temporal evolution of $\ell_{\perp}(t)$ for a set of four different seeds (data points) of wave patterns. The length scale $\ell_{\perp}(t)$ roughly grows linearly with time and saturates at $\xi_{\varphi}\equiv\ell_{\perp}(t\rightarrow\infty)=L$. \textbf{(b)} Temporal evolution of $\ell_{\perp}(t)$ for a set of six different seeds (data points) of cluster lane patterns. The length scale $\ell_{\perp}(t)$ roughly grows linearly with time and saturates at $\xi_{\varphi}\equiv\ell_{\perp}(t\rightarrow\infty)\approx0.4\,L$. Red, solid lines in \textbf{(a)} and \textbf{(b)}: averaged time evolution of $\ell_{\perp}$.
}
\label{fig:coarsening_dynamics}
\end{figure}

Having accessed the frequency of the stationary patterns of phase-segregated clusters and waves, in this section we address the dynamics of the coarsening process for both patterns. 
For classical phase-separation of fluid mixtures in the droplet-regime (asymmetric quench,~\cite{Bray_Review_1994}), coarsening of structures  is driven by Ostwald-ripening~\cite{Lifshitz_Slyozov_61} and Brownian-coalescence~\cite{Binder_Staufer_76}. Since both mechanisms rely on diffusion of either droplet material or droplets, the corresponding  growth laws exhibit identical scaling as a function of time, i.e.\ the characteristic size of structures (clusters or droplets) $\ell(t)\propto t^{1/3}$~\cite{Bray_Review_1994}.  
Interestingly, the same coefficient is reported in a model  of self-propelled particles exhibiting run-and-tumble dynamics on a lattice~\cite{Thompson_lattice_model_PP_2011}
and a similar (but slightly smaller) coarsening exponent has been found in actively propelled hard spheres~\cite{Redner_scaling_simulation_2013,Stenhammar_scaling_simulation_2013} coinciding with theoretical predictions~\cite{Stenhammar_theo_pred_scaling_2013}.
 Here, we examine the question whether there is dynamical scaling for the phase-segregated, polar-moving cluster and wave-patterns.
 
 To this end, the characteristic size of a structure $\ell(t)$ must be defined: 
 We scan over the lattice along $x$ while fixing $y$, and vice versa, and determine the corresponding probability distribution $p_{x/y}(l)$ of connected areas $l$ for which $\rho/\rho_0>1$, i.e.\ the density is larger than the homogenous density $\rho_0$. The characteristic 
 length for scanning over $x$ or $y$, respectively, is then defined as the first moment of the probability distribution,
 \begin{equation}\label{eq:char_length}
 	\ell_{x/y}(t)= \int \text{d} l \, p_{x/y}(l;t) \, l.
 \end{equation} 
  However, in contrast to a classical phase-separation in fluid mixtures~\cite{Bray_Review_1994} or actively propelled hard spheres~\cite{Redner_scaling_simulation_2013,Stenhammar_scaling_simulation_2013}, the emerging phase-segregated patterns discussed in this manuscript are not isotropic. 
  Consequently, the characteristic length has to be determined with respect to the mean polarity, i.e.\ transversal and longitudinal to it. 
  In the  following we analyze the characteristic length transversal to the stationary  polarity $\ell_\perp(t)$, whose long time asymptote gives the lateral correlation length of the pattern, i.e. $\ell_{\perp}(t\rightarrow\infty)=\xi_{\varphi}$. Speaking in terms of Eq.~\eqref{eq:char_length}, the transversal characteristic length $\ell_\perp(t)$ is equal to  $\ell_{x/y}(t)$ when the respective pattern moves along the $y/x$-axis.
 The corresponding results for wave-patterns and cluster patterns are shown in Fig.~\ref{fig:coarsening_dynamics}.
 Analyzing several random initial conditions, 
the characteristic transversal length follows roughly 
$\ell_\perp(t)\propto t^1$ for both, wave and cluster lane pattern morphologies. This linear scaling is compatible with previous results from agent-based simulations \cite{Chate_long}. While $\ell_\perp(t)$ coarsens up to the system size $L$ in case of wave patterns [Fig.~\ref{fig:coarsening_dynamics}(a)], it saturates significantly below $L$ for cluster patterns [Fig.~\ref{fig:coarsening_dynamics}(b)]. Moreover, the pronounced asymmetric fluctuations of the characteristic length in time for the cluster patterns correspond to configurations where anti-parallel moving clusters are positioned directly transversal to each other. 

Here, we would like to derive a simplified picture to explain the observed linear scaling for the domain size $\ell_\perp(t)\propto t^1$. In the following we are led by the coarsening dynamics in classical binary mixtures~\cite{Bray_Review_1994}. Say all polar high-density domains  move ballistically with a velocity $\sim1$ (measured in units of $v_0$) 
and are of characteristic size $\sim\ell_\perp$. Moreover, we assume that the total surface coverage of high-density, polar domains is approximately conserved during the coarsening process, and, especially, during each coalescence event of two polar domains. This implies that the characteristic time-scale between two coalescence events scales according to $\tau\sim\ell_\perp$. 
On average, coalescence causes the radius polar droplets to increase by a factor of $\sqrt{2}$ per coalescence event in 2D, where we assume colliding domains of approximately equal size.
We then infer the following scaling picture for relative change of the polar domain size
\begin{equation}
	\frac{1}{\ell_\perp} \frac{\text{d}}{\text{d}t} \ell_\perp\sim\frac{\Delta \ln{\ell_\perp}}{\tau}\sim\frac{\ln{2}}{2\,\tau} \sim \frac{1}{\ell_\perp},
\end{equation}
implying that $\ell_\perp(t)\sim t^1$.

Our analysis for the coarsening process within the phase-segregated regime of the phase diagram [see Fig.~\ref{fig:phase_diagram_cuts}(c)] is of course limited by the system sizes investigated (here: $L=150$). We hope that future studies involving larger system sizes confirm our observation of dynamical scaling in the coarsening dynamics of polar-moving phase-segregated morphologies.

\section{\label{sec:Outlook}Conclusion}

In this work, we have developed a general numerical framework to solve the Boltzmann equation for two dimensional systems of self-propelled particles, which we referred to as \SNAKE~(\underline{s}olving \underline{n}umerically \underline{a}ctive \underline{k}inetic \underline{e}quations). Since we have chosen to solve the Boltzmann equation in real space, our algorithm is applicable even far beyond the onset of macroscopic order, so long as the basic assumptions underlying the Boltzmann equation itself (e.g. molecular chaos) are fulfilled. Specifically, our numerical framework lends itself to the implementation of arbitrary boundary geometries, and we discussed how boundary conditions can be formulated directly in terms of microscopic particle-wall interactions.

We applied our algorithm to study an archetypal two-dimensional model system of actively propelled particles with binary, polar interactions, as previously proposed by Bertin et al. \cite{Bertin_short,Bertin_long}, and which we refer to as ``BDG model''. The BDG model is a kinetic model which is inspired by Vicsek's agent based model \cite{Vicsek}. It considers a collection of spherical, self-propelled particles moving at a constant speed $v$ such that each particle's state can be captured by a two-dimensional position vector and an angular variable characterizing the current particle orientation, i.e. direction of its velocity vector. In the absence of noise, particle orientations change in response to binary particle collisions which cause the velocity vectors of both collision partners to align along the average orientation of both particles prior to their collision. In addition, noise effects are assumed to influence both the free motion of particles between subsequent collisions (``self-diffusion''), as well as the alignment process of the binary collisions themselves.

We showed that our algorithm is fully consistent with previous analytical results on the BDG model itself \cite{Bertin_long}, as well as with the most pertinent results from extensive simulation studies on the Vicsek model \cite{Chate_long} at the onset of collective motion. In particular, our numerical results corroborate the first order nature of the phase transition toward collective motion and reproduces hysteresis effects and, therefore, phase coexistence between a spatially homogeneous, isotropic state and the emergence of polar ordered, density segregated patterns; cf. \cite{Chate_long,Ihle:2013jg}. In the framework of our numerical investigations of the BDG model, we have established two novel key insights. 

First, we have studied the emergence of macroscopic collective motion, including parameter regions well beyond the onset of macroscopic order. Our results convey a comprehensive picture, highlighting the core features of the transition between a fully isotropic, disordered state at low densities / high noise values, and a long-range ordered, spatially homogeneous state at high densities / low noise values. We found that this ordering transition bears strong structural similarities to liquid-gas transitions in equilibrium systems. There, if the density of a gas is isothermally increased beyond the binodal, the gas phase becomes metastable and, therefore, susceptible to large magnitude density fluctuations. If present, such large fluctuations lead to nucleation and growth of high density liquid droplets and eventually lead to phase separation into spatially coexisting liquid and gas phases. At still larger densities, the gas phase becomes unstable and spinodal decomposition occurs. In this case, arbitrarily small density fluctuations are amplified and the equilibrium system immediately separates into spatially coexisting low-density gas and high-density liquid phases. The mole fractions of these coexisting phases depend linearly on the system's specific volume such that the liquid phase completely takes over for large enough densities. The same line of arguments applies to the opposite case, where the density of a liquid system is being decreased. Intriguingly, we find the exact same overall behavior in the context of the ordering transition in the BDG model; cf. Fig. \ref{fig:HD_hysteresis}(b). At constant noise values (``temperature''), we identified binodal ($\rho_i$) and spinodal ($\rho_t$) density thresholds, beyond which the isotropic ``gas'' phase becomes metastable (susceptible to large enough density fluctuations) and unstable, respectively. Beyond the spinodal density threshold, $\rho_t$, arbitrarily small density fluctuations give rise to the formation of polar droplets, whose size grows linearly over time until a macroscopic, density segregated state has developed. Within this regime, well ordered high density waves or clusters (``liquid phase'') are moving on a low-density, isotropic background (``gas phase''). Similar to the equilibrium case, the surface fractions of both phases linearly depend on the system's overall density and the ``gas phase'' gets extinct at high enough overall densities. Conversely, any spatially homogeneous polar ordered system at high densities becomes metastable when the overall density falls below the second binodal density threshold, $\rho_h$, and spinodal decomposition occurs upon further reduction of the overall density below the second spinodal density, $\rho_{dh}$. 
An analogous discussion, highlighting the similarity between flocking and liquid-gas transitions, can be found in a recent work on a lattice model of active Ising spins \cite{2013PhRvL.111g8101S}. Despite the fact that both, the BDG model and the active Ising spin model of Ref. \cite{2013PhRvL.111g8101S} belong to different symmetry classes, the qualitative features of the flocking transition to spatially homogeneous polar order remain virtually unchanged. This, in turn, strongly suggests a universal character of the flocking transition to homogeneous polar order across different modeling classes and underlying symmetries.

Second, our results indicate the emergence of intriguing spatial structures after spinodal decomposition, which are genuinely distinct from wave-like patterns of infinite lateral extent. Inside the parameter regime where density segregated patterns are found, we observed a novel type of patterns, we referred to as ``\emph{cluster lane patterns}''. These patterns consist of parallel lanes of polar clusters, traveling in \emph{opposite} directions. We gave a detailed discussion of the physical mechanisms underlying the stability of this limit-cycle solution of the Boltzmann equation in terms of the transverse cross-lane transport of particles. Most importantly, these cluster lane patterns differ in two essential respects from the familiar solitary wave patterns, with which they seem to coexist in the DSP regime of the parameter space: First of all, while the solitary wave solution of the Boltzmann equation reflects the polar symmetry of the constituent particles and their mutual interactions, the cluster lane solution of the Boltzmann equation renders the overall symmetry of the ordered state nematic. Secondly, while the solitary wave solution gives rise to purely diffusive particle currents in the lateral direction which ``die out'' in the limit of long times, the cluster lane solution is hallmarked by periodically recurring propagative lateral particle transport modes which persist even for asymptotically long times.
Regarding their potentially spectacular consequences, we hope that future research will further clarify on the role of these intriguing patterns.

To conclude, the Boltzmann equation provides a convenient platform to implement particle-level descriptions of active model systems, and to assess the ensuing spatiotemporal dynamics on macroscopic scales. While this latter step is prohibitively difficult to achieve in analytical treatments of the Boltzmann equation, our numerical framework allows to study the macroscopic properties of active systems even far beyond the onset of collective order. \SNAKE, therefore, provides the basis to study a countless variety of binary collision models with arbitrary interaction symmetries, which can be achieved by appropriately (re)defining the collision and diffusion operators; cf. sections \ref{sec:sd_operator} and \ref{sec:col_operator}. More general modeling classes can be constructed by redefining the convection operator, section \ref{sec:ConvectiveOperator}, which, for instance, allows for straightforward realization of density dependent motilities \cite{Farrell:2012vf}. In particular, the kinetic framework discussed in this work is easily extended to a multi species description of active matter \cite{Weber_NJP_2013}. Therefore, \SNAKE~grants direct access to a wealth of new and exciting questions dealing with the spatiotemporal dynamics of actively propelled systems in conjunction with chemical reactions or in the context of mutually competing species both, in free space and in arbitrary confinement geometries.

\begin{acknowledgments}
We thank Hugues Chat\'e and Eric Bertin for stimulating discussions.
This work was supported by the Deutsche Forschungsgemeinschaft in the framework of the SFB 863 ``Forces in Biomolecular Systems'' (Project No. B2), and the German Excellence Initiatives via the program ``NanoSystems Initiative Munich (NIM)''.
\end{acknowledgments}

\appendix

\section{\label{app:waves}Wave patterns revisited}

\setlength{\tabcolsep}{2em}
\begin{figure*}[t]
\begin{tabular}{ccc}
\includegraphics{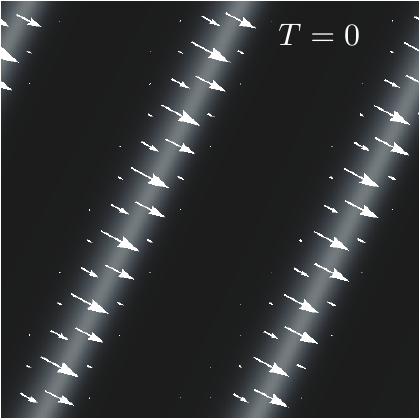} & \includegraphics{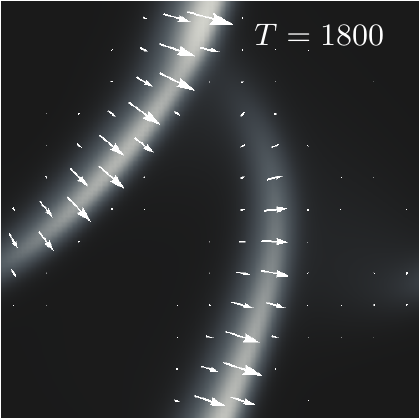} & \includegraphics{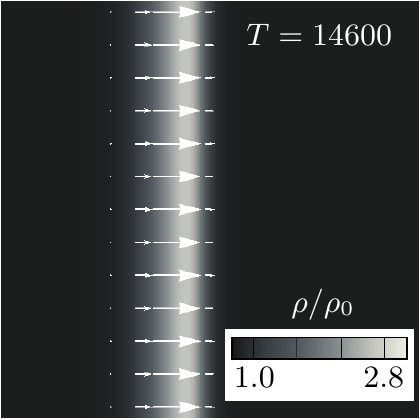}
\end{tabular}
\caption{Illustration of the bending instability in the vicinity of the transition DSP $\rightarrow$ IHP. Snapshots from left to right taken at different times $T$. The left panel indicates the initial extended wave pattern, which becomes unstable toward a bending instability (middle panel); the right panel shows the reinforced stable wave pattern which is contracted along the lateral dimension as compared to the initial pattern.}
\label{fig:bending_instability}
\end{figure*}

\setlength{\tabcolsep}{-3.4em}
\begin{figure}
\centering
\begin{tabular}{cc}
\includegraphics{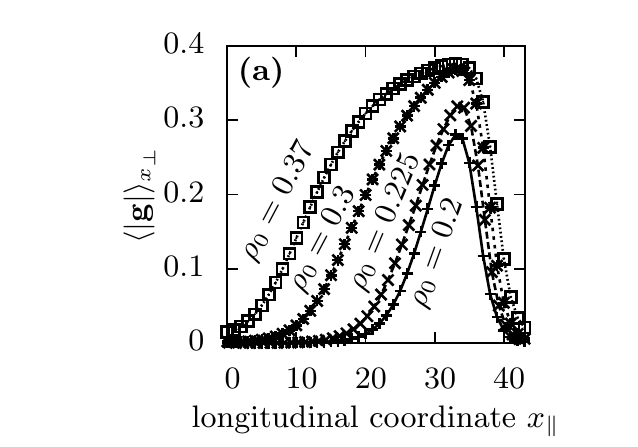} & \includegraphics{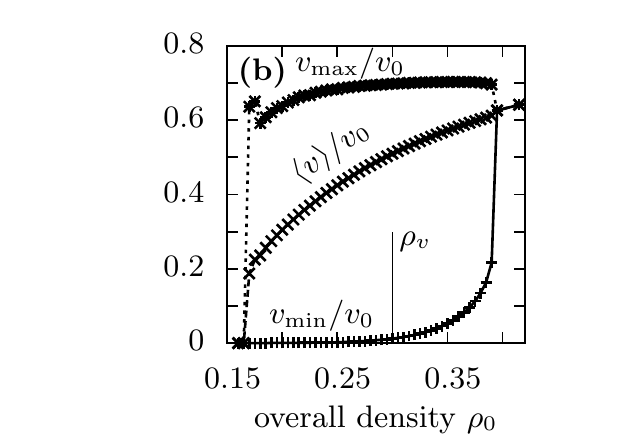}
\end{tabular}
\caption{
Momentum and velocity for wave-like patterns within the DSP regime. \textbf{(a)} Momentum profiles corresponding to the density profiles displayed in Fig.~\ref{fig:density_segregation}(a). Velocity profiles morphologically resemble the corresponding density profiles and exhibit the same increase in asymmetry with increasing density. \textbf{(b)} Characteristic velocities inside the LD phase ($v_{\text{min}}$), and the HD phase ($v_{\text{max}}$), respectively. The middle curve gives the spatially averaged velocity $\langle v\rangle$. All velocities are measured in units of $v_0$. The velocity characterizing the LD phase, $v_{\text{min}}$, attains finite values for $\rho_0\gtrsim\rho_v\approx0.3$.
}
\label{fig:momentum_profiles}
\end{figure}

In this appendix, we briefly comment on momentum and velocity fields for wave-like patterns inside the DSP regime, as well as on their bending instability at the DSP $\rightarrow$ IHP transition.

In section \ref{sec:DSP} we investigated the range of stability of wave-like patterns, as the overall density of the system is gradually decreased or increased. In both cases, our results suggested a first order transition toward a spatially homogeneous isotropic or polar state, respectively. Thereby, the former case, where the system undergoes a phase transition toward a spatially homogeneous, isotropic state upon decreasing the density below the threshold value $\rho_i$, is accompanied by a ``reinforcement'' of the wave pattern just before the DSP $\rightarrow$ IHP phase transition occurs; cf. highlighted data points in Fig.~\ref{fig:density_segregation}(b) and corresponding data points in Fig.~\ref{fig:momentum_profiles}(b). This reinforcement effect can be explained by the finite system size and is illustrated by the sequence of snapshots displayed in Fig.~\ref{fig:bending_instability}. There, the initial state of the system corresponds to a wave pattern of relatively large lateral extent, which is wrapped around the torus (representing the system's domain $\Omega$) twice [Fig.~\ref{fig:bending_instability}, left panel]. As the overall density is decreased toward the transition threshold, $\rho_0\searrow\rho_i$, this laterally elongated pattern eventually becomes unstable toward a bending instability [Fig.~\ref{fig:bending_instability}, middle panel]. However, on a periodic domain, this instability does not necessarily destroy the wave pattern, but rather can lead to a reorientation of the wave so as to contract its lateral extension [Fig.~\ref{fig:bending_instability}, right panel]. This lateral contraction, in turn, leads to a reinforcement of the wave pattern, with both, particle density and momentum across the contracted pattern being increased as compared to the initially elongated wave pattern; cf. Figs. \ref{fig:density_segregation}(b) and \ref{fig:momentum_profiles}(b). Note, however, that this reinforcement mechanism is an artifact of the finite system size and has no analog in infinite systems.

We conclude this appendix by briefly reviewing the characteristics of the momentum and velocity fields for wave patterns inside the DSP regime. To this end, we have recorded the waves' momentum profiles at various overall density $\rho_0$ inside the DSP regime, Fig.~\ref{fig:momentum_profiles}(a), and the characteristic particle velocities inside the LD phase ($v_{\text{min}}$) and HD phase ($v_{\text{max}}$), together with the system's spatially averaged velocity, $\langle v \rangle$, as a function of $\rho_0$, Fig.~\ref{fig:momentum_profiles}(b). The waves' momentum profiles closely resemble their respective density profiles both, with respect to their asymmetric shape and their growing asymmetry with increasing density. As for the characteristic velocities inside the DSP regime, we observe an approximately linear increase of the spatially averaged velocity $\langle v \rangle$ after an apparently discontinuous jump at the DSP $\rightarrow$ IHP transition. The characteristic velocity inside the HD phase remains virtually constant over the whole span of the DSP regime. In contrast, the characteristic velocity inside the LD phase is vanishingly small only up to densities $\rho_0\lesssim0.3\equiv\rho_v$, in which case the wave patterns can, in fact, be characterized as high density bands moving on an isotropic low-density sea of particles. For higher densities inside the DSP regime (i.e. $\rho_0<\rho_h$), wave patterns persist but with increasing longitudinal extensions such that the waves' tails start to infiltrate the LD phase. As a result of this infiltration, the characteristic velocity inside the LD phase is being increased to a density dependent, finite value, although the corresponding characteristic density of the LD phase remains virtually unchanged at a subcritical value, $\rho_{\text{min}}<\rho_t$; cf. Figs. \ref{fig:density_segregation}(a,b). One possible explanation for this observation of finite velocities inside the LD phase could be as follows: Since $\rho_{\text{min}}<\rho_t$, the dynamics of the locally averaged momentum inside the LD phase is such that $|\vec g_{\text{min}}|$ (and thus $v_{\text{min}}$) decays exponentially. Above a certain threshold width of the density waves (i.e. for $\rho_0>\rho_v$), however, the time scale between subsequent arrivals of wave maxima becomes comparable to the characteristic decay time scale for local momenta / velocities at $\rho=\rho_{\text{min}}$. Hence, in the stationary limit, the LD phase attains non-vanishing average particle velocities, despite the fact that local conditions in the LD phase are such that polar order tends to be destroyed.

\section{\label{sec:cluster_lane_orientations}Cluster lane orientations}

\setlength{\tabcolsep}{2em}
\begin{figure*}
\centering
\begin{tabular}{ccc}
\includegraphics{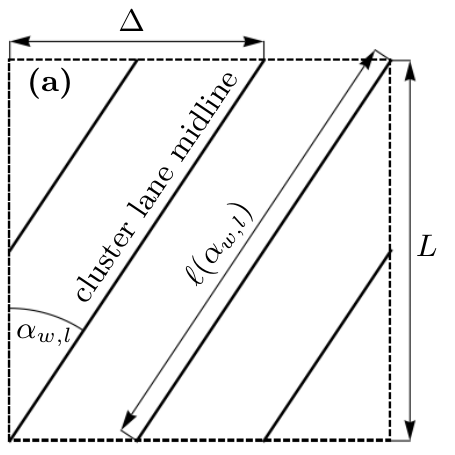} & \includegraphics{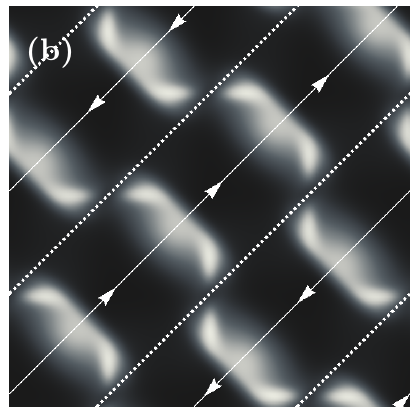} & \includegraphics{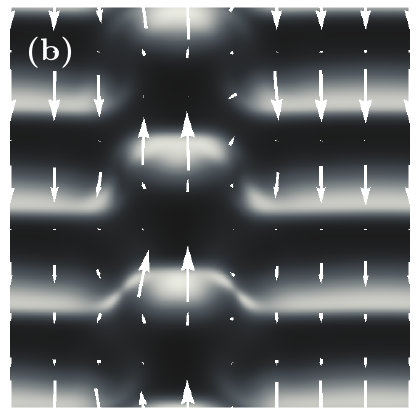}
\end{tabular}
\caption{Cluster lane geometry in periodic systems. \textbf{(a)} Schematic of the position of a cluster lane's midline for \emph{winding number} $w=2$ and \emph{length parameter} $l=3$. \textbf{(b)} Illustration of a cluster lane pattern with orientation $\alpha=\alpha_{1,1}=\pi/4$, observed in a system of linear extent $L=125$. Cluster lanes are indicated by dashed white lines; cluster lanes' midlines indicated by thin solid lines. Direction of motion indicated by arrows. \textbf{(c)} Asymmetric cluster lane pattern observed in a system of linear extent $L=150$. Parameters for (b) and (c): $\sigma=\sigma_0=0.5$; $\rho=0.25$.}
\label{fig:cluster_lane_geometry}
\end{figure*}

\begin{figure}[t]
\includegraphics{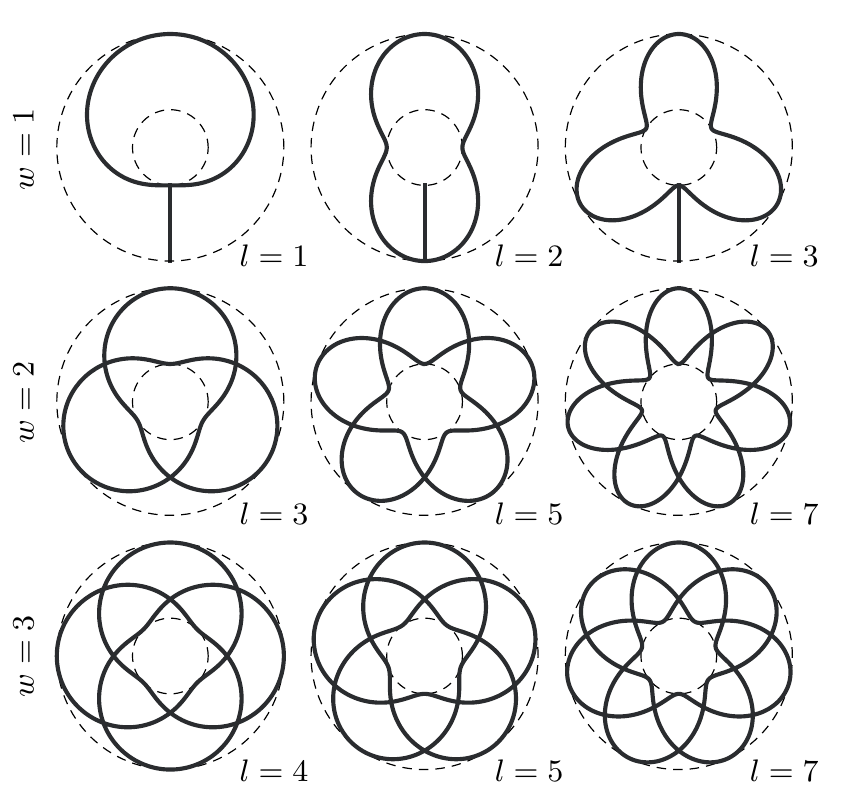}
\caption{Illustration of the \emph{winding number} $w$ and the \emph{length parameter} $l$; cf. Eq.~\eqref{eq:closure_condition}. Each plot indicates the projection of the cluster lane's midline onto the torus' equatorial plane (the system's extent in the equatorial plane is indicated by a pair of dashed, concentric circles). The \emph{winding number} $w$ (rows) gives the number of full revolutions about the torus' midpoint in the equatorial plane. The special case $w=0$ is indicated by the straight solid lines in the top row. The \emph{length parameter} $l$ (columns) gives the total length of the closed contour representing the cluster lane's midline.}
\label{fig:winding_number}
\end{figure}

Here, we will give a brief account of possible cluster lane orientations in systems with periodic boundary conditions. In particular, we will illustrate that the observation rates for cluster lane patterns do strongly depend on cluster lane orientations, and, therefore, explain why only a small number of typical cluster lane orientations is expected to be observed in simulations with finite system sizes. 

For the sake of greater clarity, we will consider systems with a $1:1$ aspect ratio. Further, we will use the one-dimensional ``midline'' of each cluster lane [i.e. the zero of the time averaged particle current inside the \emph{cluster zone}; cf. Fig.~\ref{fig:cluster_particle_currents}(a)] to represent the entire (two-dimensional) lane. Since clusters moving on different lanes (and antiparallel directions) must not collide, each cluster lane on the system's periodic domain $\Omega$ must be closed. This can expressed by requiring
\begin{equation}
\label{eq:closure_condition}
\frac{\Delta}{L}=\frac{w}{l}\leq1,\qquad w\in\mathbb N,\, l\in\mathbb N\setminus\{0\},
\end{equation}
where $\Delta$ denote the distance along the $x$-axis between subsequent intersections of the cluster lane's midline with the lines $y=0$ and $y=L$, and where $w$ and $l$ are coprime integers; cf. Fig.~\ref{fig:cluster_lane_geometry}(a). Since all values of $\Delta$ with $\Delta_1/L=L/\Delta_2$ are equivalent upon interchanging $x$- and $y$-axis, we can assert $\Delta\leq1$ without loss of generality.

In Eq.~\eqref{eq:closure_condition} the integers $w$ and $l$ characterize the topological and geometrical properties of the cluster lane, as is shown in Fig.~\ref{fig:cluster_lane_geometry}(a). Specifically, $w$ gives the cluster lane's \emph{winding number} in the equatorial plane of the torus representing the periodic domain $\Omega$; cf. Fig.~\ref{fig:winding_number}. The \emph{length parameter} $l$ measures the total length of the cluster lane in units of the (orientation dependent) length parameter $\ell(\alpha_{w,l})=L/\cos(\alpha_{w,l})$, where
\begin{equation}
\label{eq:cluster_orientation}
\alpha_{w,l}=\arctan(w/l)
\end{equation}
denotes the cluster lane's orientation with respect to the $y$-axis.

According to Eq.~\eqref{eq:closure_condition}, the ratio $\Delta/L$ can take on arbitrary rational values. At first sight, this does not seem to imply a severe restriction, since the set of rational numbers is a dense subset of the real numbers. In other words, arbitrary cluster lane orientations $\alpha$ are ``infinitely close'' to meeting the requirement Eq.~\eqref{eq:closure_condition}. While this is true in the hydrodynamic limit, $L\rightarrow\infty$, finite size effects considerably restrict the number of possible cluster lane orientations $\alpha_{w,l}$. To see this, note that the area $s_{\text{cluster}}(\alpha)$ each cluster lane of orientation $\alpha$ covers on $\Omega$ is given by ($\Lambda$: cluster lane's transversal extension)
\begin{equation}
\label{eq:space_req_cluster1}
s_{\text{cluster}}(\alpha_{w,l})=l\cdot\ell(\alpha_{w,l})\cdot\Lambda.
\end{equation}
Since the total system size is finite, $||\Omega||=L^2$, and since the minimum number of cluster lanes is $2$ (one cluster lane alone corresponds to a wave pattern), we have
\begin{equation}
\label{eq:space_req_cluster2}
s_{\text{cluster}}(\alpha_{w,l})\leq\frac{L^2}{2}.
\end{equation}
Therefore, noting that the cluster lane width $\Lambda$ is proportional to the transverse correlation length $\xi_{\varphi}$ of the polar order parameter, and that $w\leq l$, we get
\begin{equation}
w\leq l\leq\frac{L^2}{2\,\ell(\alpha_{w,l})\,\Lambda}\propto\frac{L}{\Lambda}\propto\frac{L}{\xi_{\varphi}}.
\end{equation}

For the system sizes used in our numerical computations, we found $\xi_{\varphi}\sim L=\mathcal{O}(10^2)$. We, therefore, observe cluster lane patterns characterized by $0\leq w\leq l \sim1$, i.e. $w/l\in\{0,1\}$, amounting to cluster lane orientations $\alpha_{w,l}\in\{\pi/2,\,\pi/4\}$; cf. Figs. \ref{fig:cluster_lane_geometry}(b,c). This corresponds to the cluster lane geometries indicated in the top left plot in Fig.~\ref{fig:winding_number}. Visual inspection of Fig.~\ref{fig:winding_number} illustrates the rapidly increasing space requirements for cluster lane patterns with increasing values of $w$ and $l$. These space requirements, in turn, directly affect the competition between cluster lane and wave patterns during the pattern selection process:
Although Eq.~\eqref{eq:closure_condition} has to be fulfilled for any line, representing the lateral extension of a wave pattern, as well, the resulting space requirements [as captured by Eqs.~\eqref{eq:space_req_cluster1} and \eqref{eq:space_req_cluster2}] are far less restrictive for waves. More precisely, a single wave band can attain any orientation $\alpha_{w,l}$ such that 
\begin{equation}
\label{eq:wave_space_requirement}
s_{\text{wave}}(\alpha_{w,l})=l\cdot\ell(\alpha_{w,l})\cdot\xi_{\parallel}\lesssim L^2,
\end{equation}
where $\xi_{\parallel}$ denotes the longitudinal extension of the wave. Since $\xi_{\parallel}\ll\Lambda=\mathcal{O}(\xi_{\varphi})$, Eq. \eqref{eq:wave_space_requirement} is therefore compatible with a much larger set of orientations $\alpha_{w,l}$ than Eqs.~\eqref{eq:space_req_cluster1} and \eqref{eq:space_req_cluster2} (for finite system sizes $L$). Finite system sizes $L$, therefore, imply a bias in favor of the formation of wave patterns.

\section{\label{sec:numerical_protocol_hd_hyst}Numerical protocol to measure the high density hysteresis loop}

In this appendix, we briefly explain the numerical protocol used to obtain the data points shown in Fig. \ref{fig:HD_hysteresis} (open symbols). Special care must be taken to properly assess the location of the density scale $\rho_{dh}$, where spatially homogeneous polar ordered states become linearly unstable toward the emergence of density segregated patterns. At the point where the homogeneous system turns linearly unstable, the development of spatial heterogeneities out of a virtually homogeneous state is extremely slow. 
As a sensitive means to detect the buildup of spatial heterogeneities, we introduce the maximum absolute deviation,
\begin{equation}
\Delta_f\equiv\max_{\alpha,n}\bigr\{ |f_n^{\alpha}-\overline{f}_n|\bigr\}
\end{equation}
for the \emph{density matrix} $f_n^{\alpha}$, where
\begin{equation}
\overline{f}_n\equiv\langle f_n^{\alpha}\rangle_{\alpha}.
\end{equation}
This maximum absolute deviation approaches zero for spatially homogeneous systems and attains finite values [typically of order $\mathcal{O}(10^{-1})$] for states inside the DSP regime.

To estimate the value of $\rho_{dh}$, we proceed as follows. We start from a spatially homogeneous base state, equilibrated at $\rho_0=0.415$ with $\Delta_f=\mathcal{O}(10^{-8})$. We then quench the overall density $\rho_0$ to successively lower values and let the system evolve for a time span of at least $T=30000\,\lambda^{-1}$ \footnote{Systems with densities $0.3<\rho_0\leq0.34$ had to be run for even longer time spans up to $T=160000\,\lambda^{-1}$ in order to observe the formation of wave patterns. For $\rho_0\in\{0.335,0.34\}$, we observed a monotonic but extremely slow increase in $\Delta_f$. To keep computation times at a tolerable level, we therefore perturbed the system by setting $f_n^{\alpha}\rightarrow (0.995+0.01\zeta)f_n^{\alpha}$ at $T=30000\,\lambda^{-1}$ (random variable $\zeta$ uniformly distributed on $[0,1]$)}. During this time, we record the evolution of the quantity $\Delta_f$. For all data points in Fig. \ref{fig:HD_hysteresis} with $V_{\text{hd}}/||\Omega||=1$, the maximum absolute deviation $\Delta_f$ is a monotonically decreasing function of time, and the system approaches a spatially homogeneous state. Conversely, in systems being represented by data points with $V_{\text{hd}}/||\Omega||<1$, $\Delta_f$ increases such that the spatially homogeneous initial states eventually give rise to density segregated patterns.

As mentioned earlier, for density quenches down to final densities $\rho_0\lesssim\rho_{dh}$ in the vicinity of $\rho_{dh}$, the actual buildup of spatial heterogeneities happens on a very large time scale. In actual computations, this time scale turns out to be much slower than the local ``equilibration'' of the density matrix $f_n^{\alpha}$ toward the new density level $\rho_0$. We, therefore, do not expect that the results shown in Fig. \ref{fig:HD_hysteresis} should depend on whether they are obtained from a ``quasi-static'' protocol, or a density quench protocol as discussed here.


%

\end{document}